\documentclass[aps,floatfix,superscriptaddress,nofootinbib,,tightenlines,longbibliography,preprintnumbers,11pt]{revtex4-1}
\usepackage{babel}
\usepackage{natbib}
\usepackage{amssymb,amsthm,amsfonts,braket,amsmath,graphicx,color,dsfont, mathrsfs}
\usepackage{array}
\newcolumntype{C}[1]{>{\centering\let\newline\\\arraybackslash\hspace{2pt}}m{#1}}
\usepackage{bm}
\usepackage[export]{adjustbox}
\usepackage{enumerate}
\usepackage{adjustbox}
\usepackage{xfrac}
\usepackage{tikz}
\usepackage{quantikz}
\usepackage{url}
\usepackage{xcolor}
\usepackage{multirow}
\usepackage{rotating,booktabs}
\usepackage[verbose]{placeins}
\usepackage{ragged2e}
\usepackage[titletoc]{appendix}
\usepackage{slashed}
\usepackage{comment}
\usepackage{outlines}
\usepackage[colorlinks=true,citecolor=blue,linkcolor=blue
]{hyperref}
\usepackage{soul}
\usepackage{enumitem}
\usepackage{float}
\setlist[enumerate]{label*=\arabic*.}
\begin{document}

\dimen\footins=5\baselineskip\relax

\preprint{UMD-PP-025-02, IQuS@UW-21-102
}

\title{
Quantum computation of hadron scattering in a lattice gauge theory
}

\author{Zohreh Davoudi}
\email{davoudi@umd.edu}
\affiliation{Department of Physics and Maryland Center for Fundamental Physics (MCFP), University of Maryland, University of Maryland, College Park, MD 20742 USA}
\affiliation{Joint Center for Quantum Information and Computer Science (QuICS), National Institute of Standards and Technology (NIST) and University of Maryland, College Park, MD 20742 USA}
\affiliation{The NSF Institute for Robust Quantum Simulation, University of Maryland, College Park, Maryland 20742, USA}
\affiliation{National Quantum Laboratory (QLab), 
\\
University of Maryland, College Park, MD 20742 USA}

\author{Chung-Chun Hsieh}
\email{cchsieh@umd.edu}
\affiliation{Department of Physics and Maryland Center for Fundamental Physics (MCFP), University of Maryland, University of Maryland, College Park, MD 20742 USA}
\affiliation{Joint Center for Quantum Information and Computer Science (QuICS), National Institute of Standards and Technology (NIST) and University of Maryland, College Park, MD 20742 USA}
\affiliation{The NSF Institute for Robust Quantum Simulation, University of Maryland, College Park, Maryland 20742, USA}

\author{Saurabh V. Kadam}
\email{ksaurabh@uw.edu}
\thanks{corresponding author.}
\affiliation{
InQubator for Quantum Simulation (IQuS) and Department of Physics, University of Washington, Seattle, WA 98195, USA}

\begin{abstract}
We present a digital quantum computation of two-hadron scattering in a $Z_2$ lattice gauge theory in 1+1 dimensions. We prepare well-separated single-particle wave packets with desired momentum-space wavefunctions, and simulate their collision through digitized time evolution. Multiple hadronic wave packets can be produced using the efficient, systematically improvable algorithm of this work, achieving high fidelity with the target initial state.
Specifically, employing a trapped-ion quantum computer (\texttt{IonQ Forte}), we prepare up to three meson wave packets using 11 and 27 system qubits, and simulate collision dynamics of two meson wave packets for the smaller system.
Results for local observables are consistent with numerical simulations at early times, but decoherence effects limit evolution into long times. We demonstrate the critical role of high-fidelity initial states for precision measurements of state-sensitive observables, such as $S$-matrix elements.
Our work establishes the potential of quantum computers in simulating hadron-scattering processes in strongly interacting gauge theories.
\end{abstract}

\maketitle


\tableofcontents

\section{Introduction
\label{sec: Introduction}
}
A plethora of scattering experiments continue to serve as an excellent testing ground for verifying theoretical predictions for the structure of hadrons and nuclei~\cite{Accardi:2012qut, Anderle:2021wcy, Deshpande:2005wd,Achenbach:2023pba}, searching for new particles and interactions beyond the Standard Model~\cite{Narain:2022qud, Bruning:2012zz,Shiltsev:2019rfl,ParticleDataGroup:2022pth,Kadan:2024ezm,DUNE:2020fgq,SNO:2015wyx,Takeuchi:2022dfj}, and investigating phases of matter in and out of equilibrium~\cite{csernai1994introduction,Ludlam:2005cfp,Andronic:2017pug,Berges:2020fwq,Lovato:2022vgq}.
Many of these experiments involve atomic nuclei, which are composed of nucleons that interact via the strong force. Quantum chromodynamics (QCD) is the quantum theory of the strong force, which is a non-Abelian gauge theory that describes the interactions between the constituents of nucleons, quarks and gluons~\cite{Gross:2022hyw}. QCD becomes strongly interacting at low energies, hence non-perturbative physics takes effect. Furthermore, the theory exhibits confinement at low energies: the interacting degrees of freedom are bound states of quarks and gluons. Classical computers~\cite{Joo:2019byq}, in conjunction with the lattice-QCD method~\cite{Wilson:1974sk}, have enabled first-principles non-perturbative studies of QCD observables~\cite{FlavourLatticeAveragingGroupFLAG:2024oxs,Davoudi:2022bnl,USQCD:2022mmc}.
In the lattice-QCD approach, the spacetime is discretized and time is rendered imaginary. Quantities are evaluated in a path-integral formalism using Monte Carlo methods, where in most scenarios, the action leads to a well-defined probability distribution~\cite{Montvay:1994cy,Rothe:1992nt,Gattringer:2010zz}. 
Lattice QCD is most suitable for calculating static observables since it uses an Euclidean-time action. Nonetheless, it can also be employed to extract dynamical observables like scattering and decay amplitudes~\cite{Briceno:2014tqa,Briceno:2017max,Hansen:2019nir,Davoudi:2020ngi}, thanks to theoretical developments that utilize the finite-volume nature of lattice-QCD calculations~\cite{Luscher:1986pf, Luscher:1990ux}.
However, the application of such methods is limited to low-energy and low-inelasticity processes. Currently, first-principles predictions for general real-time scattering processes, e.g., those relevant for descriptions of high-energy particle collisions and the early universe, remain beyond the reach of the state-of-the-art lattice-QCD calculations~\cite{Gattringer:2016kco,Nagata:2021ugx}. 

An alternative to the Euclidean-time Lagrangian formulation of lattice gauge theories (LGTs) is the Hamiltonian formulation~\cite{Kogut:1974ag}.
In this formulation, one constructs the Hamiltonian, and the associated Hilbert space, of a gauge theory on a discretized space but with a continuous time. Real-time processes are then computed from time evolution of states. Interest in the Hamiltonian LGTs has revived in the recent years thanks to novel computational paradigms that offer polynomial, instead of exponential, scaling of the computational resources with respect to system's volume.
One paradigm is the tensor-network methods that achieve polynomial scaling by systematically controlling the entanglement generation in the state when possible, using suitable tensor-network ansatzes~\cite{Banuls:2019rao,Meurice:2020pxc, Banuls:2022vxp, Cirac:2020obd}. These calculations are gaining attention for studying real-time dynamics in LGTs~\cite{Magnifico:2020bqt, Silvi:2014pta, Tagliacozzo:2014bta,Rico:2013qya,Buyens:2015tea, Pichler:2015yqa, Buyens:2016hhu,Kuhn:2015zqa,Mathew:2025fim,Banuls:2025wiq}, including scattering problems~\cite{Rigobello:2021fxw, Belyansky:2023rgh, Papaefstathiou:2024zsu, Jha:2024jan}.
Nonetheless, utilizing this tool for simulating high-dimensional models and  non-Abelian gauge theories remains limited~\cite{Magnifico:2024eiy,Kelman:2024gjt, Yosprakob:2023jgl}. The second novel computational paradigm uses quantum computers, leveraging the principles of quantum superposition and entanglement~\cite{Feynman:1981tf, Simon:2006oay, Lloyd:1996aai, Preskill:2021apy}, and offering exponential scaling in its (more conventional) building blocks of qubits~\cite{Nielsen:2012yss} or its rapidly emerging alternatives of qudits~\cite{Chi:2023hmy,Nikolaeva:2023tuk}, fermions~\cite{Bravyi:2000vfj}, or bosons~\cite{Chabaud:2022hoq}.
LGTs subsequently can be simulated by mapping their Hilbert space to that of the constituent degrees of freedom of a quantum computer~\cite{Dalmonte:2016alw,Banuls:2019bmf,Klco:2021lap,Bauer:2022hpo,Bauer:2023qgm,Halimeh:2023lid,DiMeglio:2023nsa,Beck:2023xhh}. One way to simulate a Hamiltonian time evolution is by expressing the time-evolution operator in terms of a universal set of single-qubit gates, and two-qubit entangling gates.
This approach, which is known as digital quantum simulation, is quantum-hardware agnostic, and will be the basis of this work.

In a pioneering work by Jordan, Lee, and Preskill~\cite{Jordan:2012xnu,Jordan:2011ci}, an algorithm for digital quantum simulation of scattering was proposed for a scalar field theory with quartic interactions. This algorithm adopts the $S$-matrix approach in the scattering theory, where the ``in'' (``out'') state is taken to be single-particle wave packets long in the past (future) and far separated in space. The initial scattering state of well-separated single-particle wave packets is prepared from the adiabatic evolution of a known state. The known state is taken to be two well-separated single-particle wave packets in the non-interacting theory, and adiabatic evolution amounts to gradually turning on the interaction strength. Once the wave packets in the interacting theory are prepared, the state is evolved under the full Hamiltonian until they collide. Finally, measurements are performed by simulating detectors in the late-time regions.
Adiabatic state preparation has, nonetheless, several practical and theoretical drawbacks.
For example, it requires an overhead to compensate for wave-packet broadening during the evolution~\cite{Jordan:2012xnu}.
Furthermore, adiabatic state preparation becomes computationally infeasible when a phase transition occurs on the evolution path~\cite{Farhi:2000ikn,Surace:2024bht}---a situation that incurs, e.g., when one tries to prepare the ground state of the confined phase of a model from that of the deconfined phase.
In the context of quantum field theories, improvements to the adiabatic state preparation have been proposed by using more efficient bases~\cite{Barata:2020jtq}, by traversing efficient paths within the adiabatic trajectory~\cite{Cohen:2023dll}, or by combining variational~\cite{Klco:2018kyo, Atas:2021ext, Fromm:2023npm, Crippa:2024hso} and adiabatic~\cite{Chakraborty:2020uhf,DAnna:2024mmz} methods in an scalable manner~\cite{Farrell:2023fgd,Farrell:2024fit,Li:2024lrl}.
Recently, algorithms for wave-packet preparation and two-particle scattering in (1+1)D scalar field theory have been demonstrated on quantum computers~\cite{Ingoldby:2025bdb,Zemlevskiy:2024vxt}, albeit limited to highly truncated fields and short-time evolutions, due to decoherence. Finally, a first attempt at observing inelastic scattering in a (1+1)D Ising field theory was reported in Ref.~\cite{Farrell:2025nkx}.

Proposals exist for preparing the initial interacting wave packets by circumventing the adiabatic state preparation~\cite{Liu:2021otn,Turco:2023rmx,Turco:2025jot,Kreshchuk:2023btr,Hite:2025pvb}, including extraction of the scattering observables from appropriate asymptotic matrix elements directly~\cite{Li:2023kex,Briceno:2023xcm,Briceno:2020rar,Gustafson:2021imb,Peruzzo:2013bzg,McClean:2015vup,Tilly:2021jem, Sharma:2023bqu,Ciavarella:2020vqm}. 
Nonetheless, these approaches are either limited to massive excitations in the process or are challenging to apply to studies of high-energy particle collisions, including subsequent hadronization, fragmentation, and thermalization processes.
Another method that requires no adiabatic interpolation and, instead, creates the wave packets directly in the interacting theory is that proposed in Ref.~\cite{Rigobello:2021fxw} in a (1+1)D U(1) LGT using tensor-network methods, and used in Ref.~\cite{Chai:2023qpq} to study fermion-antifermion scattering.
We recently provided an improvement to this approach in Ref.~\cite{Davoudi:2024wyv} by using an efficient hybrid classical-quantum variational algorithm that creates wave packets of bound excitations in confining gauge theories in (1+1)D. We further demonstrated the algorithm's viability in the noisy intermediate-scale quantum (NISQ) era by implementing it on a trapped-ion quantum computer.

In the present work, we further optimize our state-preparation method by parameterizing a position-space rather than a momentum-space form, which proves to be more efficient. We employ this algorithm together with a time-evolution algorithm to simulate scattering in a (1+1)D $Z_2$ LGT with dynamical fermions.
This LGT exhibits non-perturbative features, such as confinement. It, thus, serves as a testing ground for simulating the complex theory of QCD, and at the same time, it may provide insights into the physics of confinement~\cite{Magnifico:2019ulp,Charles:2023zbl,Mildenberger:2022jqr,Irmejs:2022gwv,Frank:2019jzv,Borla:2019chl,Borla:2020upy,Das:2021iwa,Samajdar:2022mtt,Konig:2019trl,Yarloo:2018zvu}. First, we show a resource-efficient method for preparing the initial state of well-separated single-particle wave packets with tunable parameters, like their spatial width and central momentum. 
The new approach of this work allows for various controlled approximations to the target state, and in the worst case, relies on an ansatz that uses only a polynomial number of parameters in the system size. Moreover, for theories with finite correlation length, the algorithm's cost scales polynomially with the correlation length rather than with the system size. We demonstrate our state-preparation algorithm by preparing multiple single-particle wave packets on an \texttt{IonQ Forte} quantum computer~\cite{Chen:2023erd} for two different lattice sizes, and compare it against the very high fidelity (and thus quantum-resource intensive) state obtained using the noiseless emulator, observing good agreement.

Importantly in this work, we proceed with evolving the two wave packets till they collide.
Various methods have been proposed for simulating time evolution in the recent years~\cite{Low:2016znh,Kirby:2021ajp,Chen:2021zva,Kalev:2020xvu,Low:2018pte}. Trotter product expansion of the unitary time-evolution operator, nonetheless, was found to be the most resource-efficient algorithm for our NISQ-era application.
Our goal is to investigate the type of scattering observables accessible using near-term NISQ-era devices.
We first quantum compute on the same hardware the real-time evolution of local observables. 
We show that these observables agree reasonably well with the ideal result at early times, and employing a symmetry-based noise mitigation during post-processing has little effect.
We further show, with the aid of the noisy emulator provided by \texttt{IonQ}, that additional noise-mitigation circuits could be implemented to retrieve the signal for observables that are more deviated from their idea values.
Finally, we highlight the importance of preparing high-fidelity initial states by comparing the probability of returning to the initial state for different degrees of initial-state infidelities.
This quantity is ultimately related to the scattering $S$-matrix, and for unstable single-particle states, it plays a role in calculating the decay width~\cite{Yamada:2023wqp, Izrailev_2006,Ciavarella:2020vqm}.
We find that such a non-local observable, as expected, is highly sensitive to the approximations made in the initial state.

The paper is organized as follows.
In Sec.~\ref{sec: Theoretical formalism}, we provide the necessary theoretical background for constructing the single-particle wave packets and the ansatz used to create them.
Section~\ref{sec: Quantum Algorithm and Circuit Design} describes the different components of the quantum algorithm employed to prepare and evolve the wave packets.
Section~\ref{sec: Results} presents both the quantum hardware and emulator results, along with the noise-mitigation methods used.
Finally, we conclude, and provide an outlook, in Sec.~\ref{sec: Conclusion and Outlook}. Several appendices supplement our results and analyses.

\section{Theoretical formalism
\label{sec: Theoretical formalism}
}

This section contains the required theoretical background for implementing the scattering protocol of this work on a quantum hardware. First, an efficient  formulation of a $Z_2$ LGT in (1+1)D is derived in Sec.~\ref{subsec: Z2 theory EGF and MGF}, then an ansatz for building the lowest-energy particle excitation in each momentum sector is developed and optimized in Sec.~\ref{subsec: Ansatz form}. Finally, numerical evidence for the performance of the optimized ansatz, and its accuracy in regimes of arbitrary coupling and mass, are presented in Sec.~\ref{subsec: ansatz continuum limit numerical analysis}.

\subsection{$Z_2$ LGT in (1+1)D with dynamical matter
\label{subsec: Z2 theory EGF and MGF}
}

Consider a $Z_2$ LGT coupled to a single flavor of fermions on a spatial lattice with a periodic boundary condition (PBC).
The spatial lattice of $N$ lattice points is denoted by  $\Gamma = \big \{0,a,\cdots,(N-1)a \big \}$, where $a$ is the lattice spacing.
Matter contents in this theory are staggered fermions~\cite{Kogut:1974ag, Banks:1975gq}: fermions and antifermions live on the even and odd sites of the lattice, respectively, implying that $N$ must be an even number, and the staggered lattice corresponds to a physical system with $N_{P} \coloneq N/2$ lattice sites.
The fermionic creation (annihilation) operator at site $n$ is denoted by $\xi^\dagger_n$ ($\xi_n$).
The (spin-$\frac{1}{2}$) hardcore bosons hosted on the lattice links are the gauge-boson degrees of freedom, represented by the gauge-link operator $\tilde{\sigma}^{\mathbf{x}}_n$ for the link connecting sites $n$ and $n+a$, and its non-commuting conjugate variable, the electric-field operator, $\tilde{\sigma}^{\mathbf{z}}_n$.

The local Hilbert space at site $n$ is spanned by $\ket{f_n}$ with $f_n=0,1$, the staggered-fermion occupation. In the Dirac-sea picture, $f_n=0$ $(1)$ at even lattice sites corresponds to the absence (presence) of a particle, while at odd lattice sites it corresponds to the presence (absence) of an antiparticle. Note that, $\xi_n \ket{0_n} = \xi^\dagger_n \ket{1_n} = 0$, $\xi_n \ket{1_n} = \ket{0_n}$, and $\xi^\dagger_n \ket{0_n} = \ket{1_n},$ $\forall n \in \Gamma$. 
The local Hilbert space at the link originating from site $n$ in the electric-field basis is spanned by $\ket{s_n}$ with $s_n=\uparrow_n,\downarrow_n$, the two spin projections of a hardcore boson along the z-axis at site $n$.
The gauge-link and electric-field operators in this basis are thus given by the Pauli $\mathbf{x}$ and $\mathbf{z}$ operators $\tilde{\sigma}^{\mathbf{x}}_n=\ket{\uparrow_n}\bra{\downarrow_n}+\ket{\downarrow_n}\bra{\uparrow_n}$ and $\tilde{\sigma}^{\mathbf{z}}_n=\ket{\uparrow_n}\bra{\uparrow_n}-\ket{\downarrow_n}\bra{\downarrow_n}$, respectively.

A generic basis state in this Hilbert space is given by
\begin{equation}
    \ket{\psi} = \ket{f_0,f_a,\cdots,f_{(N-1)a}}\otimes\ket{s_0,s_a,\cdots,s_{(N-1)a}}.
    \label{eq: generic state in KS}
\end{equation}
Only a portion of this Hilbert space is physically relevant because the physical states $\ket{\psi}_{\text{phys}}$ must satisfy the local Gauss's laws
\begin{align}
    & G_n\ket{\psi}_{\text{phys}} = \ket{\psi}_{\text{phys}},~\forall n,\label{eq: Gauss's law condition}
\end{align}
where
\begin{equation}
    G_n = \tilde{\sigma}^{\mathbf{z}}_n \tilde{\sigma}^{\mathbf{z}}_{n-1}e^{i\pi\bar{Q}_n}.
    \label{eq: Gauss's law operator def}
\end{equation}
Here,
\begin{equation}
    \bar{Q}_n = 
    \xi^\dagger_n\xi_n -\frac{1-(-1)^{n/a}}{2}.
    \label{eq: Q at n in KS}
\end{equation}
such that $\bar{Q}_n = 1\,(-1)$ when a fermion (antifermion) is present at $n$, and $\bar{Q}_n = 0$ otherwise.

The system's Hamiltonian $H$ is given by
\begin{equation}
    aH = \frac{1}{2} \sum_{n \in \Gamma}{\left( \xi_{n}^\dagger \tilde{\sigma}^{\mathbf{x}}_n \xi_{n+a} + \text{H.c.}\right)} + 
    am_f\sum_{n \in \Gamma}{(-1)^{n/a}  \xi_{n}^\dagger\xi_n }+
    a\epsilon \sum_{n \in \Gamma} \tilde{\sigma}^{\mathbf{z}}_n,
    \label{eq: Z2 Ham KS}
\end{equation}
where $m_f \geq 0 $ is the fermion mass and $\epsilon$ is the strength of the electric-field Hamiltonian.
Here, $\xi_{Na}$ is identified with $\xi_0$ due to the PBC.
From here onward, we set $a=1$, hence quantities are expressed in units of lattice spacing.
The PBC yields lattice translational invariance and allows for specifying well-defined momentum quantum numbers. 

The Gauss's law operator in Eq.~\eqref{eq: Gauss's law operator def} commutes with the Hamiltonian in Eq.~\eqref{eq: Z2 Ham KS}.
The Hamiltonian also commutes with a global operator $Q$ given by the total fermionic occupation in the lattice:
\begin{equation}
    Q = \sum_{n=0}^{N-1} \xi^\dagger_n\xi_n.
    \label{eq: global quantum number def}
\end{equation}
Following our previous work~\cite{Davoudi:2024wyv}, we restrict this study to the subspace of the Hilbert space spanned by states with $Q=N_P$.
A special state in this subspace, called the strong-coupling vacuum (SCV), is the ground state of the Hamiltonian in the limit of $m_f,\epsilon \gg 1$, and is given by
\begin{equation}
    \ket{\Omega}_\text{SCV}= \ket{0,1,\cdots,0,1} \otimes \ket{s,s,\cdots,s,s}.
    \label{eq: SCV in EGF def}
\end{equation}
Here, $s=\,\uparrow$ $(\downarrow)$ for $\epsilon < 0$ ($\epsilon > 0$).

The gauge degrees of freedom in (1+1)D LGTs are not dynamical in nature, and are often rotated away when open boundary condition (OBC) are in place. In this case, only the fermionic-matter degrees of freedom survive but at the expense of yielding a non-local Hamiltonian.
Such an elimination of the gauge bosons is achieved by transforming the fermionic fields as
\begin{subequations}
\label{eq: fermion transformation for MGF}
\begin{align}
    \xi_n \to \psi_n = \left(\prod_{i=0}^{n-1} \tilde{\sigma}^{\mathbf{x}}_i \right) \xi_n,\\
    \xi_n^\dagger \to \psi_n^\dagger = \left(\prod_{i=0}^{n-1} \tilde{\sigma}^{\mathbf{x}}_i \right) \xi_n^\dagger.
\end{align}
\end{subequations}
With the PBC, the gauge degrees of freedom cannot be completely rotated away. However, they are reduced to a single spin degree of freedom on one link, which can be seen by examining the Hamiltonian in Eq.~\eqref{eq: Z2 Ham KS} under the field re-definitions in Eq.~\eqref{eq: fermion transformation for MGF}:
\begin{equation}
    H = \frac{1}{2} \sum_{n=0}^{N-2}{\left( \psi_{n}^\dagger \psi_{n+1} + \text{H.c.}\right)} + \frac{1}{2}\left({ \psi_{N-1}^\dagger\, \widetilde{\Sigma}^{\mathbf{x}}\, \psi_{0} + \text{H.c.}}\right) + m_f\sum_{n \in \Gamma}{(-1)^{n}  \psi_{n}^\dagger \psi_n }+\epsilon \sum_{n \in \Gamma} \widetilde{\Sigma}^{\mathbf{z}}_n.
    \label{eq: Z2 Ham MGF}
\end{equation}
Here,
\begin{equation}
    \widetilde{\Sigma}^{\mathbf{x}}  \coloneq \prod_{i=0}^{N-1} \tilde{\sigma}^{\mathbf{x}}_n,
    \label{eq: MGF Sigma X operator}
\end{equation}
and the second term in Eq. \eqref{eq: Z2 Ham MGF} resulted from the PBCs.
The electric-field operator $\widetilde{\Sigma}^{\mathbf{z}}_n$ is evaluated using the Gauss's law in Eq.~\eqref{eq: Gauss's law condition}:
\begin{equation}
    \widetilde{\Sigma}^{\mathbf{z}}_n \coloneq \tilde\sigma^\textbf{z}_n = e^{i\pi\sum_{m=0}^{n} \bar{Q}_m} \,\tilde{\sigma}^{\textbf{z}}_{N-1},
    \label{eq: Sigma Z at n in MGF}
\end{equation}
where $\bar{Q}_n$ is obtained by substituting Eq.~\eqref{eq: fermion transformation for MGF} in Eq.~\eqref{eq: Q at n in KS}:
\begin{equation}
    \bar{Q}_n = 
    \psi^\dagger_n\psi_n -\frac{1-(-1)^n}{2}.
    \label{eq: Q at n in MGF}
\end{equation}
The factor $\sum_{m=0}^{n} \bar{Q}_m$ can be either an even or an odd integer, resulting in $\widetilde{\Sigma}^{\mathbf{z}}_n = \tilde{\sigma}^{\textbf{z}}_{N-1}$ or $\widetilde{\Sigma}^{\mathbf{z}}_n = -\tilde{\sigma}^{\textbf{z}}_{N-1}$, respectively. (The restriction $\sum_{n=0}^{N-1} \xi^\dagger_n\xi_n = \sum_{n=0}^{N-1} \psi^\dagger_n\psi_n = N_P$ implies $\sum_{n=0}^{N-1} \bar{Q}_n = 0$, yielding $\widetilde{\Sigma}^{\mathbf{z}}_{N-1}=\tilde{\sigma}^{\textbf{z}}_{N-1}$.) As a result, the conjugate-variable field to the electric field $\pm \tilde \sigma_{N-1}^\textbf{z}$ at the last link is $\pm \tilde \sigma_{N-1}^\textbf{x}$. 
The action of $\widetilde{\Sigma}^{\mathbf{x}}$ on $\ket{\psi}$ defined in Eq.~\eqref{eq: generic state in KS} is, therefore, equivalent to the action of $\pm \tilde{\sigma}^{\textbf{x}}_{N-1}$ on $\ket{s_{N-1}}$. (Either of these options yields the same spectrum so we choose a positive sign.) In other words, we use the convention that the action of the transformed gauge-link operator $\widetilde{\Sigma}^{\mathbf{x}}$ is now restricted to the link connecting lattice sites $N-1$ and $0$.\footnote{While the Hamiltonian in Eq.~\eqref{eq: Z2 Ham MGF} appears to distinguish the last link from the rest of the links, lattice translational symmetry of the original formulation is still preserved in this merely gauge-transformed formulation.}

One can then reduce the bosonic Hilbert space to the local Hilbert space at the last link, and a generic state in this Hilbert space is now given by
\begin{equation}
    \ket{\psi}_{\rm phys} = \ket{f_0,f_1,\cdots,f_{N-1}}\otimes\ket{s_{N-1}}.
    \label{eq: generic state in MGF}
\end{equation}
The $\psi_n$ and $\psi^\dagger_n$ field operators act on the fermionic Hilbert space in the same manner as $\xi_n$ and $\xi^\dagger_n$ operators act on the state in Eq.~\eqref{eq: generic state in KS}.
The action of the gauge-link and the electric-field operator at site $n$ is restricted to the reduced bosonic Hilbert space. The electric Hamiltonian, nonetheless, has become non-local, where the non-locality originates from the dependence of $\widetilde{\Sigma}^{\mathbf{z}}_n$ on the fermion-number occupations up to site $n$.
Non-locality is the cost to pay to render all states physical in this Hilbert space (there is no local Gauss's law constraint anymore in the new formulation).
When subjected to the restricted subspace that satisfies $Q=N_P$, the dimension of the Hilbert space is $2\times(2N_P)!/(N_P!)^2$, and the SCV state is given by
\begin{equation}
    \ket{\Omega}_\text{SCV}= \ket{0,1,\cdots,0,1} \otimes \ket{s_{N-1}},
    \label{eq: SCV in MGF def}
\end{equation}
with $s_{N-1}=\uparrow$ $(\downarrow)$ for $\epsilon < 0$ ($\epsilon > 0$).

We refer to the Hamiltonian in Eq.~\eqref{eq: Z2 Ham KS} and the Hilbert space spanned by states in Eq.~\eqref{eq: generic state in KS} as the explicit gauge-link formulation (EGF), and to the Hamiltonian in Eq.~\eqref{eq: Z2 Ham MGF} and its corresponding Hilbert space spanned by states in Eq.~\eqref{eq: generic state in MGF} as the minimal gauge-link formulation (MGF) of a $Z_2$ LGT in (1+1)D with PBCs.\footnote{Such alternative formulations have been studied in the past for different LGTs with PBCs, e.g., in Ref.~\cite{Nagele:2018egu, Zache:2020qny,Zache:2018cqq}.}
Quantum simulation of MGF requires fewer qubits than the EGF, as can be seen from mapping of fermionic and bosonic degrees of freedom to qubits.
The fermions in this work are mapped to qubits via the Jordan-Wigner transformation
\begin{subequations}
\label{eq: Jordan-Wigner}
\begin{align}
        \psi_n^\dagger = \left(\displaystyle \prod_{m=0}^{n-1}{\sigma^{\textbf{z}}_m}\right) \sigma_n^{-}, \label{eq: Jordan-Wigner for psi dagger}\\
        \psi_n = \left(\displaystyle \prod_{m=0}^{n-1}{\sigma^{\textbf{z}}_m}\right) \sigma_n^+,
    \label{eq: Jordan-Wigner for psi}
\end{align}
\end{subequations}
to satisfy the anti-commuting nature of fermionic field operators. Here, $\sigma^{\textbf{z}}_i$ is the Pauli-$\mathbf{z}$ operator and $\sigma^{+}_i$ $(\sigma^{-}_i)$ is the Pauli-raising (lowering) operator acting on the $i^{\rm th}$ qubit.
Hence, in both formulation, $N$ qubits are needed to represent the fermionic Hilbert space.
The hardcore bosons are mapped to qubits directly with the $s$ state denoting the corresponding qubit eigenstate in the z-basis.
The qubit-resource savings is evident: to express the bosonic Hilbert space, the MGF needs only one qubit while the EGF requires $N$ qubits.
In this work, we use the MGF for simulating scattering in a $Z_2$ LGT, which requires $N+1$ qubits for mapping the Hilbert space of $N$ ($N_P$) staggered (physical) lattice sites.

The qubit Hamiltonian used for the unitary time evolution is obtained by substituting Eq.~\eqref{eq: Jordan-Wigner} in Eq.~\eqref{eq: Z2 Ham MGF}, resulting in
\begin{equation}
    H = H^h + H^m + H^\epsilon,
    \label{eq: Z2 Ham MGF JW in Hh Hm He}
\end{equation}
with
\begin{subequations}
    \begin{align}
        H^h &= \frac{1}{4} \sum_{n=0}^{N-2}{\left( \sigma^{\textbf{x}}_{n}\sigma^{\textbf{x}}_{n+1} + \sigma^{\textbf{y}}_{n}\sigma^{\textbf{y}}_{n+1}\right)}
        +\frac{\alpha_N}{4}\left({ \sigma^{\textbf{x}}_{N-1}\tilde{\sigma}^{\textbf{x}}_{N-1}\sigma^{\textbf{x}}_{0}+ \sigma^{\textbf{y}}_{N-1}\tilde{\sigma}^{\textbf{x}}_{N-1}\sigma^{\textbf{y}}_{0}}\right),\phantom{\left(\prod_{j=0}^n \sigma^{\textbf{z}}_{j}\right)} \label{eq: Hh def}\\
        H^m &= \frac{m_f}{2}\sum_{n \in \Gamma}{(-1)^{n+1}  \sigma^{\textbf{z}}_{n}},\phantom{\left(\prod_{j=0}^n \sigma^{\textbf{z}}_{j}\right)}\label{eq: Hm def}\\
        H^\epsilon &= \epsilon \,\tilde{\sigma}^{\textbf{z}}_{N-1} + \epsilon \sum_{n=0}^{N-2}\gamma_n
        \tilde{\sigma}^{\textbf{z}}_{N-1}\left(\prod_{j=0}^n \sigma^{\textbf{z}}_{j}\right). \label{eq: Hepsilon def}
    \end{align}
    \label{eq: Hh Hm Hepsilon defs set}
\end{subequations}
Here, $H^h$ obtains the interaction (or the hopping) energy, $H^m$ obtains the fermion-mass energy, and $H^\epsilon$ obtains the electric-filed energy.
The factor $\alpha_N \coloneq (-1)^{N_P+1}$ in Eq.~\eqref{eq: Hh def} arises from pulling to the right a string of ${\sigma^{\textbf{z}}}$ Paulis that span the entire lattice and evaluating it for the subspace considered here, i.e., $Q=N_P$. The factor $\gamma_n$ in Eq.~\eqref{eq: Hepsilon def} is $\gamma_n \coloneq e^{-i\pi\sum_{m=0}^{n} (1-(-1)^m)/2}=i^{n}$ for even $n$ and $i^{n+1}$ for odd $n$. The first term in the same equation results from $e^{i\pi\sum_{m=0}^{N-1} \bar{Q}_m}=1$ in the $Q=N_P$ sector.
Note that, the operators acting on the gauge-boson qubit are distinguished from the fermion qubits with an overhead tilde notation.

In the next subsection, we discuss how to build the single-particle wave-packet state that comprises the initial scattering state.

\subsection{Ansatz for meson creation operators in the interacting theory
\label{subsec: Ansatz form}
}

To simulate two-meson scattering in a (1+1)D $Z_2$ LGT on a quantum computer, one first needs to prepare an initial state made up of two single-particle wave packets, ideally separated far away from each other.
The form of the single-particle wave packets $\ket{\Psi}$ follows our earlier work in Ref.~\cite{Davoudi:2024wyv}. The state $\ket{\Psi}$ is defined as a collection of the lowest-energy single-particle states, $\ket{k}$, in each momentum sector $k$, smeared by the wave-packet profile $\Psi(k)$:
\begin{equation}
    \ket{\Psi} = \sum_{k\in\widetilde{\Gamma}
    } {\Psi(k) \ket{k}}, \label{eq: WP def}
\end{equation}
where $\widetilde{\Gamma} \coloneq \frac{\pi}{N_P} \big\{-N_P,-N_P+1, \cdots, N_P-1 \big\}\cap [-\frac{\pi}{2}, \frac{\pi}{2})$ corresponds to the physical momenta on the lattice.
In Ref.~\cite{Davoudi:2024wyv}, each $\ket{k}$ was obtained by optimizing an ansatz for the creation operator $b_k^\dagger$. Explicitly, $b_k^\dagger$ excites $\ket{k}$ when acted on the interacting vacuum $\ket{\Omega}$:
\begin{equation}
    \ket{k} = b_k^\dagger\ket{\Omega},
    \label{eq: bk dagg on vaccum}
\end{equation}
with $\braket{\Omega|\Omega}=\braket{k|k}=1$.
This optimization can be done using a variational quantum eigensolver (VQE). The ansatz in Ref.~\cite{Davoudi:2024wyv} was motivated by a similar ansatz proposed earlier in Ref.~\cite{Rigobello:2021fxw} in the context of tensor-network calculations of scattering in a U(1) gauge theory.

In this work, we propose a closely-related alternative ansatz whose form is motivated by the translation symmetry of the physical lattice. It uses kinematic factors similar to the ansatz in Refs.~\cite{Rigobello:2021fxw,Davoudi:2024wyv}, which are obtained by solving the free fermionic theory on a staggered lattice.
The translation symmetry of the staggered lattice, where a unit translation on the original (physical) lattice turns into a translation by two units on the staggered sites, is incorporated in the ansatz by considering the gauge-invariant operators $\mathcal{M}_{m,n}$.
The operator $\mathcal{M}_{m,n}$ for the simple case of $m=n$ reduces to the fermion number operator $\xi^\dagger_m\xi_m$.
However, when $m\neq n$, $\mathcal{M}_{m,n}$ is given by a non-local operator called a bare-meson creation operator.
In the EGF, this operator is composed of $\xi^\dagger_m$, $\xi_n$, and a string of $\tilde{\sigma}^{\mathbf{x}}_l$ operators connecting them.
For $m<n$ and the periodic lattice, there can be two ways of constructing such an operator: $\xi^\dagger_m \,(\prod_{l=m}^{n-1}\tilde{\sigma}^{\mathbf{x}}_l) \, \xi_n$ or $\xi_n \, (\prod_{l=n}^{N-1}\tilde{\sigma}^{\mathbf{x}}_l) (\prod_{l=0}^{m-1}\tilde{\sigma}^{\mathbf{x}}_l) \, \xi^\dagger_m$.
The former is referred to as a forward-wrapped $(n-m)$-length meson creation operator and the latter a backward-wrapped $(N-n+m)$-length meson creation operator.
When $m>n$, the forward-wrapped (backward-wrapped) meson creation operators are given by similar expressions with the fermionic operators ordered accordingly, such that the forward-wrapped meson is given by $\xi^\dagger_m \, (\prod_{l=m}^{N-1}\tilde{\sigma}^{\mathbf{x}}_l) (\prod_{l=0}^{n-1}\tilde{\sigma}^{\mathbf{x}}_l) \, \xi_n$ and the backward-wrapped meson is given by $\xi_n \,(\prod_{l=n}^{m-1}\tilde{\sigma}^{\mathbf{x}}_l) \, \xi^\dagger_m$. 
We require that the ansatz only include the shorter meson depending on $m$ and $n$, and if the meson length is equal for both forward and backward wrapping, i.e., $|m-n|=N_P$, each operator is added with a coefficient $\frac{1}{\sqrt{2}}$ such that each bare meson is created with a probability of $\frac{1}{2}$.

\begin{table}[t!]
    \renewcommand{\arraystretch}{2}
    \begin{center}
        
    \begin{tabular}{|C{1cm}| C{2.2cm}||C{3.9cm}|C{1.7cm}|C{6.7cm}|}
    \hline
    \multicolumn{2}{|c||}{Case} & EGF & MGF & $\widetilde{\mathcal{M}}_{m,n}$\\
    \hline
    \multicolumn{2}{|c||}{$m=n$} & $\xi^\dagger_m\xi_m $ &  $\psi^\dagger_m\psi_m$ & $\frac{1}{2}\left(\mathds{1}-\sigma^z_m \right)$ \\
    \hline
    \multirow{2}{*}{$m<n$} & $|m - n|<N_P$ & $\xi^\dagger_m \,\left({\displaystyle \prod_{l=m}^{n-1}}\tilde{\sigma}^{\mathbf{x}}_l\right) \, \xi_n$ & $\psi^\dagger_m\psi_n$ & $\sigma^-_m \,\left({\displaystyle\prod_{l=m+1}^{n-1}\sigma^{\textbf{z}}_{l}}\right) \, \sigma^+_n$ \\
    & $|m - n|>N_P$ & $\xi_n \left({\displaystyle\prod_{l=n}^{N-1}\tilde{\sigma}^{\mathbf{x}}_l\prod_{l=0}^{m-1}}\tilde{\sigma}^{\mathbf{x}}_l\right) \xi^\dagger_m $ & $\psi_n \, \widetilde{\Sigma}^{\mathbf{x}} \, \psi^\dagger_m$ & $(-1)^{N_P}\left({\displaystyle\prod_{l=0}^{m-1}}\sigma^{\textbf{z}}_{l}\right)
    \sigma^-_m \sigma^+_n \left({\displaystyle\prod_{l=n+1}^{N-1}}\sigma^{\textbf{z}}_{l}\right) \tilde{\sigma}^{\textbf{x}}_{N-1}$\\
    \hline
    \multirow{2}{*}{$m>n$} & $|m - n|<N_P$ & $\xi_n \,\left({\displaystyle \prod_{l=n}^{m-1}}\tilde{\sigma}^{\mathbf{x}}_l\right) \, \xi^\dagger_m$  &  $\psi_n\psi^\dagger_m$ & $(-1)\, \sigma^+_n \,\left({\displaystyle\prod_{l=n+1}^{m-1}\sigma^{\textbf{z}}_{l}}\right) \, \sigma^-_m$\\
    & $|m - n|>N_P$ & $\xi^\dagger_m \, \left({\displaystyle\prod_{l=m}^{N-1}\tilde{\sigma}^{\mathbf{x}}_l \prod_{l=0}^{n-1}}\tilde{\sigma}^{\mathbf{x}}_l\right) \, \xi_n$ &  $\psi^\dagger_m \, \widetilde{\Sigma}^{\mathbf{x}} \, \psi_n$ & $(-1)^{N_P+1}\left({\displaystyle\prod_{l=0}^{n-1}}\sigma^{\textbf{z}}_{l}\right)\sigma^+_n \sigma^-_m \left({\displaystyle\prod_{l=m+1}^{N-1}}\sigma^{\textbf{z}}_{l}\right) \tilde{\sigma}^{\textbf{x}}_{N-1}$\\
    \hline
\end{tabular}
\end{center}
\caption{Definition of $\mathcal{M}_{m,n}$ for different cases of $m$ and $n$ values. 
The second and third column summarize the form of $\mathcal{M}_{m,n}$ in the EGF and MGF, respectively.
The last column denotes its form in the MGF after performing the Jordan-Wigner transformation in Eq.~\eqref{eq: Jordan-Wigner}, denoted here by $\widetilde{\mathcal{M}}_{m,n}$. This is later used for mapping the $b^\dagger_k$ operator to operations on qubits in a quantum circuit.
The forward-(backward-)wrapped meson operators are identified when the creation operator at $m$ appears to the left (right) of the annihilation operator at $n$ in the EGF and MGF columns. 
The $|m-n|=N_P$ case is not explicitly shown in the table for brevity: it can be easily obtained by adding the forward- and backward-wrapped mesons of length $N_P$, each with a coefficient $1/\sqrt{2}$.
\label{tab: bare meson operators}}
\end{table}
The $\mathcal{M}_{m,n}$ operators translate by two staggered lattice sites under one unit translation of the physical lattice, as mentioned above.
Thus, for a given bare meson length $l$, the operators $\mathcal{M}_{m,n}$ that start at even (odd) $m$ are mapped to each other via the physical translations.
Upon performing the gauge transformation in Eq.~\eqref{eq: fermion transformation for MGF}, the bare-meson creation operator reduces to a fermion hopping term between $m$ and $n$ if the meson does not contain the remnant gauge link, otherwise, the fermion hopping term includes an additional $\widetilde{\Sigma}^{\mathbf{x}}$ operator defined in Eq.~\eqref{eq: MGF Sigma X operator} (whose action reduces to $\tilde{\sigma}_{N-1}^{\textbf{x}}$ on the bosonic Hilbert space in the MGF).
A summary of $\mathcal{M}_{m,n}$ definitions in both EGF and MGF is given in Table~\ref{tab: bare meson operators}, along with their mappings to the qubit space in the MGF after the Jordan-Wigner transformations in Eq.~\eqref{eq: Jordan-Wigner}.

In this confined theory, it is expected that the operators which have a support over an extent much greater than the confinement scale will be suppressed exponentially.
Motivated by this intuition, we introduce an ansatz for $ b^\dagger_k$ that can be improved order by order, and takes the following form:
\begin{equation}
    b^\dagger_k = \frac{1}{\mathcal{N}}\sum_{j = 0}^{N-1} \eta^{(j)}_k.
    \label{eq: bk dagger complete ansatz definition}
\end{equation}
Here, $\mathcal{N}$ is a normalization factor that will be discussed later. The superscript $j$ denotes the $j^{\rm th}$ order at which only the bare-meson creation operators with length less than or equal to $j$ contribute.
The expression for $\eta^{(j)}_k$  is given by
\begin{equation}
    \eta^{(j)}_k = \sum_{m,n\in\Gamma^{(j)}_{0}} 
    e^{- \alpha_0^{(j),k}\left|{m-n}\right|^2
    }\overline{C}^k_{m,n}\mathcal{M}_{m,n} + \sum_{m,n\in\Gamma^{(j)}_{1}} 
    e^{- \alpha^{(j),k}_1\left|{m-n}\right|^2}\overline{C}^k_{m,n}\mathcal{M}_{m,n},
    \label{eq: eta j definition}
\end{equation}
where $\Gamma^{(j)}_{i}=\{m,n \in \Gamma \;|| \;|m-n|= j \;\text{and}\; m\mod 2 = i\}$.
Here, $\left|m-n\right|$ is taken to be the shortest distance between $m$ and $n$ on a periodic lattice. 
$\overline{C}^k_{m,n}$ are kinematical factors given by
\begin{equation}
    \overline{C}^k_{m,n} =\sum_{p,q \in \widetilde{\Gamma}} \delta_{k,p+q} \mathscr{C}(p,m)\mathscr{D}(q,n),
    \label{eq: C tilde definition}
\end{equation}
with the momentum sums in $p$ and $q$ running over the Brillouin zone of the staggered lattice, $\widetilde{\Gamma}$,
and 
\begin{subequations}
\begin{align} 
&\mathscr{C}(p,m)=\sqrt{\frac{m_f+\omega_p}{2\pi\omega_p}} {e^{ipm}}\left(\mathcal{P}_{m,0}+v_p\mathcal{P}_{m,1}\right),
\\
&\mathscr{D}(q,n)=\sqrt{\frac{m_f+\omega_q}{2\pi\omega_q}} {e^{iqn}}\left(-v_q\mathcal{P}_{n,0}+\mathcal{P}_{n,1}\right),
\end{align}
\label{eq: kinematic factors definition}
\end{subequations}
which are obtained by solving for the plane-wave solutions in the free fermion theory on a staggered lattice~\cite{Rigobello:2021fxw}.
Here, $\omega_k=\sqrt{m_f^2+\sin^2(k)}$ and $v_k=\frac{\sin(k)}{m_f+\omega_k}$. Moreover, $\mathcal{P}_{n,0(1)}=\frac
{1+(-1)^{n+0(1)}}{2}$ is the projection operator to the even (odd) staggered sites.
One can rewrite Eq.~\eqref{eq: eta j definition} using Eqs.~\eqref{eq: C tilde definition} and~\eqref{eq: kinematic factors definition} in terms of $\sum_{m,n \in \Gamma} \mathscr{C}(p,m) \mathscr{D}(q,n) \mathcal{M}_{m,n}$. The ansatz is, therefore, built in such a way to impart momentum $p$ and $q$ to the $m$ and $n$ end of the bare-meson creation operator $\mathcal{M}_{m,n}$.
The Kronecker delta $\delta_{k,p+q}$ then ensures that the total momentum of the composite object is the desired momentum $k$.
Finally, the exponential factors $e^{- \alpha_{0/1}^{(j),k}\left|{m-n}\right|^2}$ control the strength of contribution of each bare-meson creation operator according to their lengths, motivated by the discussions above.

At any given order $j$, the operator $b^\dagger_k$ can be approximated by $b^{(j)\dagger}_k$, which using Eqs.~\eqref{eq: bk dagger complete ansatz definition}-\eqref{eq: kinematic factors definition} has the form:
\begin{equation}
    b^{(j)\dagger}_k =\sum_{j'=0}^j \,\sum_{m,n\in\Gamma^{(j')}}
    C^{(j'),k}_{m,n} \mathcal{M}_{m,n},
    \label{eq: bk dagger as Cmn Mmn}
\end{equation}
where $\Gamma^{(j)} = \Gamma_0^{(j)}\cup\Gamma_1^{(j)}$ and
\begin{equation}
    C^{(j),k}_{m,n} = \frac{1}{\mathcal{N}} e^{- \alpha_i^{(j), k}\left|{m-n}\right|^2}\overline{C}^k_{m,n} \quad \text{for} \quad m,n \in \Gamma_i^{(j)}.
\end{equation}
We choose the normalization constant $\mathcal{N}$ at each order $j$ such that $\sum_{m,n\in\Gamma^{(j)}} |C^{(j),k}_{m,n}|^2 = 1$. 

For a given $k$, there are two parameters to be optimized at each order $j$, $\alpha_0^{(j),k}$ and $\alpha_1^{(j),k}$.
Given the natural hierarchy of operators based on their lengths, the order-by-order optimization scheme spares one from optimizing all the $\alpha_{0/1}^{(j),k}$ parameters simultaneously.
One can start the optimization process with $j=1$ that restricts the ansatz to length-1 bare-meson creation operators.\footnote{Note that, $\alpha_{0/1}^{(j=0),k}$ do not enter the optimization process since the ansatz does not depend on them when $m=n$.}
A variational minimization algorithm can then be run for parameters $\alpha_{0/1}^{(1),k}$ to obtain ${\alpha_{0/1}^{(1),k}}^*$ that minimizes the energy of the state $b^{(1)\dagger}_k\ket{\Omega}$.
To improve the overlaps with the true lowest-energy state with a  momentum $k$, one can then choose to go to the next order in $j$ and repeat the variational minimization process on $\alpha_{0/1}^{(1),k}$ and $\alpha_{0/1}^{(2),k}$. Nonetheless, the parameter space can be restricted to a narrow region around the optimized ${\alpha_{0/1}^{(1),k}}^*$ from the previous order of optimization to ease the next optimization.
The process is iterated for progressively larger $j$ values, while restricting the parameter space for $\alpha_{0/1}^{(\ell),k}$ with $\ell<j$ to a narrow range near the already optimized values ${\alpha_{0/1}^{(\ell),k}}^*$.
One can then expect to obtain the form of $b^\dagger_k$ that is very close to the true momentum creation operator of the interacting theory, which excites the $\ket{k}$ state when acted on the interacting vacuum $\ket{\Omega}$.

Once the ansatzes for operators $b^\dagger_k$ are optimized to reach a target accuracy, the list of optimized ${\alpha_{0/1}^{(\ell),k}}^*$ can be used in Eqs.~\eqref{eq: WP def} and~\eqref{eq: bk dagg on vaccum} to define an operator $b^\dagger_{\Psi}$. This operator creates the desired wave-packet state $\ket{\Psi}$ from the interacting vacuum:
\begin{equation}
    \ket{\Psi} = \left(\sum_{k\in\widetilde{\Gamma}} \Psi(k)b^{\dagger}_k\right) \ket{\Omega} \equiv b^{\dagger}_{\Psi}\ket{\Omega}.
    \label{eq: b psi dagger def}
\end{equation}
Furthermore, using the ansatzes for $b^\dagger_k$ from Eq.~\eqref{eq: bk dagger complete ansatz definition} and the Jordan-Wigner transformed definitions of $\mathcal{M}_{m,n}$, denoted by $\widetilde{\mathcal{M}}_{m,n}$ in Table~\ref{tab: bare meson operators}, the operator $b^\dagger_{\Psi}$ can be approximated by the operator $b^{(j)\dagger}_{\Psi}$ built of $b_k^{(j)\dagger}$ operators:
\begin{equation}
    b^{(j)\dagger}_{\Psi} = \sum_{j'=0}^j\,\sum_{m,n  \in \Gamma^{(j')}} C^{(j')}_{m,n} \, \widetilde{\mathcal{M}}_{m,n},
    \label{eq: b psi dagger after Jordan-Wigner}
\end{equation}
where 
\begin{equation}
    C^{(j)}_{m,n} = \sum_{k \in \tilde{\Gamma}} \Psi(k)\, C^{(j),k}_{m,n}.
    \label{eq:cmn-def}
\end{equation}
Thus, the coefficients $C^{(j)}_{m,n}$ depend on the optimized parameters ${\alpha_{0/1}^{(\ell),k}}^*$, the kinematic factors $\mathscr{C}$ and $\mathscr{D}$ defined in Eqs.~\eqref{eq: kinematic factors definition}, and the wave-packet profile $\Psi(k)$.
This form of $b^{(j)\dagger}_{\Psi} \approx b^\dagger_{\Psi}$ will be used in Sec.~\ref{sec: Results} for preparing the initial scattering state. 

\subsection{States' fidelities with the ansatz meson creation operators
\label{subsec: ansatz continuum limit numerical analysis}
}
\begin{figure}[t]
    \centering
    \includegraphics[scale = 1]{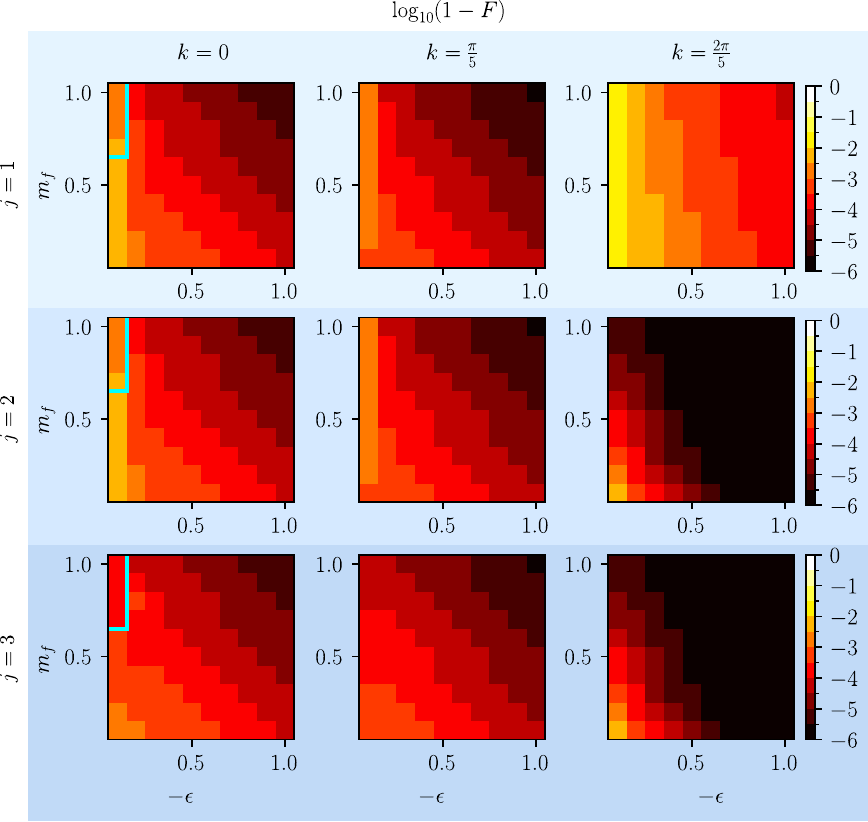}
    \caption{The $j^{\rm th}$-order ansatz infidelity $1-F$ (in logarithmic scale) as a function of Hamiltonian parameters $\epsilon$ and $m_f$ for $N=10$. Each column corresponds to a non-negative momentum $k$ in the Brillouin zone, and each row corresponds to the order of the ansatz $j$. Higher fidelity is observed for $\ket{k^{(j)}_{\rm op}}$ when larger values of $m_f, |\epsilon|$, and a higher ansatz order $j$ are used. For example, $F=0.99$ is achieved at $j=3$ in the entire parameter space for all $k$. The region to the left of the cyan contour corresponds to the existence of a low-energy non-mesonic excitation that is not captured by our mesonic ansatz.}
    \label{fig: scan_10}
\end{figure}
In this section, we demonstrate the accuracy with which the optimized ansatz represents the desired states at a given order. In small systems, the accuracy can be checked by calculating the fidelity between the momentum eigenstates obtained by exact diagonalization and the optimized ansatz.
We define the fidelity of a hadron state in the momentum sector $k$ as
\begin{equation}
    F \coloneq 
    {\left|\,{}_{\rm op}\langle k^{(j)}|k\rangle_{\rm ex}\right|}^2.
    \label{eq:fidelity}
\end{equation}
Here, $\ket{k^{(j)}}_{\rm op}$ is prepared by acting on $\ket{\Omega}_{\rm ex}$ with the optimized ansatz for $b^{(j)\dagger}_k$, where $\ket{\Omega}_{\rm ex}$ is the ground state obtained by numerically diagonalizing the Hamiltonian.
Similarly, $\ket{k}_{\rm ex}$ is the lowest-energy mesonic eigenstate with momentum $k$ obtained from the exactly diagonalized Hamiltonian.  

Figure \ref{fig: scan_10} shows the fidelity scan in a 10-(staggered) site theory over a range of Hamiltonian parameters, $m_f$ and $\epsilon$, for different $k$-momentum sectors, and at various ansatz orders.
The order-by-order optimization is conducted in the following manner: at the $j^{\rm th}$ order, a total of $2j$ parameters $\alpha_{0/1}^{(j'),k}$ with $ j'\leq j$ are optimized. The initial values of the first $2j-2$ parameters $\alpha_{0/1}^{(j'),k}$ with $j' < j$ have been set by the optimized results at the previous $({j-1})^{\rm th}$ order, and their variation are limited to a window $w=0.1\times {\rm max}(|\alpha_{0/1}^{(j'),k}|, 1)$ to aid the optimization at the $j^{\rm th}$ order.

As mentioned earlier, the $j^{\rm th}$ order ansatz consists of bare mesons with length up to $j$ in position space. 
Therefore, it is reasonable to expect that a higher-order ansatz can better capture the interacting momentum eigenstate $\ket{k}$. 
Also, as the order $j$ is increased, the ansatz is expected to improve in capturing longer-range correlations.
These expectations are in agreement with our numerical results. 
The fidelity can reach 0.98 (0.99) for all the parameter range considered if the first-(third-)order ansatz is used, and the optimization at all orders works better in the strong-coupling regime where the mass gap is larger.
We also note that the infidelities in $\ket{k^{(j)}}_{\rm op}$ are associated with contamination from the excited states in the same momentum sector $k$, since our ansatz transforms properly under translation by construction.

In the $k=0$ sector, there is one non-mesonic configuration that can be constructed by flipping the single bosonic link, which cannot be captured by the ansatz.
Such an excitation corresponds to one unit of electric flux across the periodic lattice in the EGF, and has an energy proportional to $N$. It, thus, does not generally appear in the low-energy spectrum.
Nonetheless, for sufficiently small $\epsilon$ at a given $m_f$, i.e., the area to the left of the cyan contour in Fig.~\ref{fig: scan_10}, such a non-mesonic excitation can have energies lower than the single-particle momentum states. 
In this region, $\ket{k=0}_{\rm ex}$ is defined as the first mesonic excitation in the $k=0$ sector, and the fidelity is calculated accordingly.

Compared to the results in our recent work~\cite{Davoudi:2024wyv}, which adopts a momentum-space ansatz containing only two parameters overall, the order-by-order translationally invariant position-space ansatz developed here performs better, even at the edge of the Brillouin zone (i.e., for large momenta). 
The ability to create accurate momentum eigenstates with large $k$ yields faster-moving wave packets. 
This, first of all, injects more energy into the process and leads more interesting outgoing scattering channels. Second, it reduces the total evolution time needed to bring the wave packets together in a scattering simulation; therefore, fewer Trotter steps and shallower circuits are needed for a Trotterized time evolution. 

Finally, we note that the optimized values of $\alpha_{0/1}^{(j),k}$ are not unique with different initializations of the optimization routine. However, we observe that the normalization condition for $\mathcal{N}$ renders the resulting $C^{(j)}_{m,n}$ coefficients out of these initializations approximately identical. (The $C^{(j)}_{m,n}$ coefficients determine the gate parameters in the quantum circuit.) 
It is also seen that the optimized $\alpha_{0/1}^{(j),k}$ can be different between $k$ and $-k$ momentum sectors, as opposed to being symmetric like the ansatz in Ref.~\cite{Davoudi:2024wyv}.\footnote{We have tried a symmetric ansatz with respect to $k \to -k$, which is identical to ansatz in Eq.~\eqref{eq: eta j definition} except one needs to include a factor of $i^{|m-n|}$ in both sums. We have checked that, despite restoring the symmetry, this ansatz overall returns lower fidelities than the anstaz based on Eq.~\eqref{eq: eta j definition}. We, therefore, do not consider such an ansatz any further.}
Only the positive-$k$ fidelities are plotted in Fig.~\ref{fig: scan_10} for brevity. (Similar fidelities are obtained for the negative-$k$ states).

The order-by-order ansatz is expected to work for larger systems, as demonstrated in Appendix~\ref{app: tensor network} for a 26-(staggered) site theory. 
For lattice sizes that exceed the regime of viability of exact diagonalization, we compare the ansatz against matrix-product states (MPS), which can accurately approximate the exact eigenstates in (1+1)D, upon employing a density-matrix-renormalization-group (DMRG) method.
A parameter scan similar to Fig.~\ref{fig: scan_10}, but in the 26-(staggered) site theory, is presented in Fig.~\ref{fig: scan_26}, demonstrating high fidelities achieved even with the lowest-order ansatzes.

For the remainder of the main text, to reduce the clutter, we drop the superscript $(j)$ from quantities, since the order at which they are assumed will be clear from the context.

\section{Quantum algorithm and circuit design
\label{sec: Quantum Algorithm and Circuit Design}
}
\begin{figure}[t]
    \centering
    \includegraphics[]{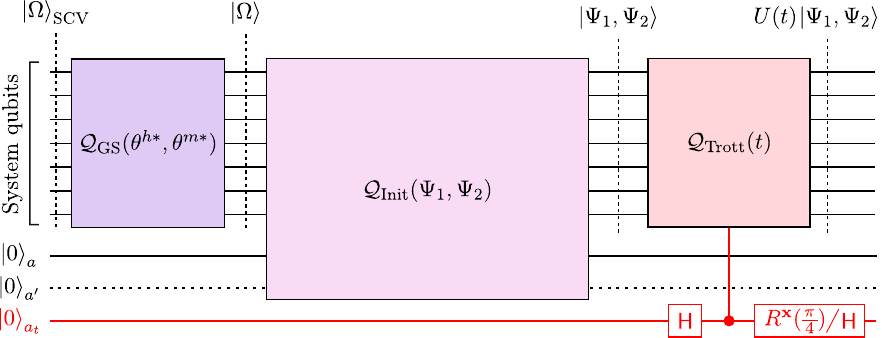}
    \caption{Shown is the circuit used in this work for simulating scattering in a (1+1)D $Z_2$ LGT in the MGF. The circuit acts on $N+1$ system qubits representing the lattice associated with $N$ staggered sites, and one or more ancilla qubits denoted by $a_{1,2,t}$. The system qubits are initialized in the SCV state $\ket{\Omega}_\text{SCV}$, as shown by the first vertical dotted line, while each ancilla starts in the $\ket{0}$ state. The circuit consists of three subcircuit modules. The vertical dotted lines indicate the state of the system qubits after the application of each module. The first module prepares the ground state, $\ket{\Omega}$, of the Hamiltonian in Eq.~\eqref{eq: Z2 Ham MGF JW in Hh Hm He} using the circuit block $\mathcal{Q}_{\rm GS}$ with parameters $\theta^{h*}$ and $\theta^{m*}$. The second module prepares the initial scattering state composed of two well-separated input wave packets, resulting in the state $\ket{\Psi_1,\Psi_2}$. It requires at least one ancilla qubit to prepare the initial state, shown here with a black solid line. Alternatively, it can also be applied using an extra ancilla qubits, shown with a dotted black line, to improve the accuracy of preparing the target state, as explained in the text. Finally, the last module, $\mathcal{Q}_{\rm Trott}$, performs the unitary time evolution $U(t)=e^{-itH}$ under the Hamiltonian $H$ in Eq.~\eqref{eq: Z2 Ham MGF JW in Hh Hm He}. The circuit components in red are to perform a Hadamard test to compute the return probability of the initial state as a function of time; they can be omitted when measuring only the expectation values of diagonal operators. The Hadamard test requires an additional ancilla and a controlled application of $\mathcal{Q}_{\rm Trott}$ with control on the ancilla. Here, $\mathsf{H}$ denotes the Hadamard gate and $R^{\textbf{x}}(\theta) \coloneq e^{-\frac{i}{2}\theta \sigma^{\textbf{x}}}$. To compute the return probability, as shown in Appendix~\ref{app: Hadamard test}, one needs to separately implement either the $R^{\textbf{x}}(\frac{\pi}{4})$ or the $\mathsf{H}$ gate as the last operation on the $a_t$ ancilla, which is denoted in the circuit by $R^{\textbf{x}}(\frac{\pi}{4})/\mathsf{H}$.
    Details of each module and their constituent circuits are discussed in Sec.~\ref{sec: Quantum Algorithm and Circuit Design}.
    }
    \label{fig: circuit flow}
\end{figure}
\begin{figure}[t]
    \centering
    \includegraphics[]{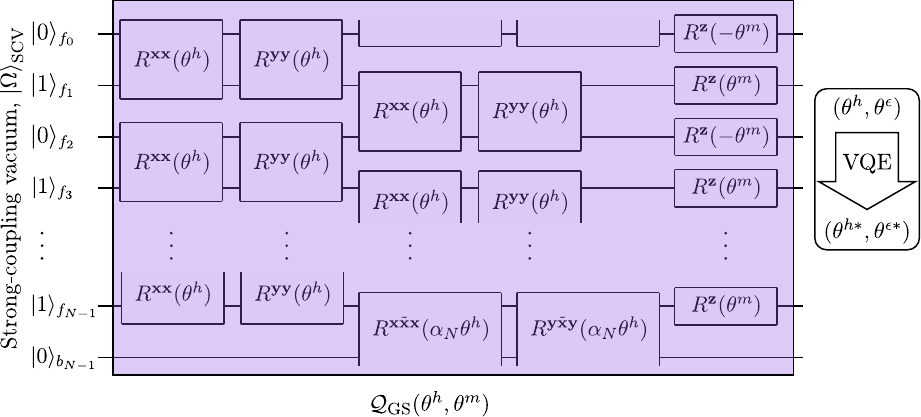}
    \caption{Shown is the circuit $\mathcal{Q}_{\rm GS}$ that prepares the interacting ground state $\ket{\Omega}$ of the Hamiltonian in Eq.~\eqref{eq: Z2 Ham MGF JW in Hh Hm He}. This circuit is parameterized by two parameters, $\theta^h$ and $\theta^m$.
    The circuit $\mathcal{Q}_{\rm GS}$ acts on the strong-coupling vacuum $\ket{\Omega}_\text{SCV}$, which is given by alternating $\ket{0}$ and $\ket{1}$ states on qubits that represent fermion lattice sites, labeled here with a subscript $f_i$ for the $i^{\rm th}$ site, and $\ket{0}$ state for the qubit that represents the bosonic link in the MGF, denoted here by the subscript $b_{N-1}$. The two-qubit gate are defined as $R^{\textbf{xx}}(\theta) \coloneq e^{-\frac{i}{2}\theta\,\sigma^{\textbf{x}}_{f_{i_1}}
    \sigma^{\textbf{x}}_{f_{i_2}}}$ and $R^{\textbf{yy}}(\theta) \coloneq e^{-\frac{i}{2}\theta \,\sigma^{\textbf{y}}_{f_{i_1}}
    \sigma^{\textbf{y}}_{f_{i_2}}}$,
    where $i_1$ and $i_2$ denote fermion lattice sites each gate involves. The three-qubit gates $R^{\textbf{x}\tilde{\textbf{x}}\textbf{x}}(\theta)$ and $R^{\textbf{y}\tilde{\textbf{x}}\textbf{y}}(\theta)$ are defined in the similar manner with the overhead tilde indicating the qubit operation on the bosonic link qubit $b_{N-1}$, and $\alpha_N$ is defined below Eq.~\eqref{eq: Hh Hm Hepsilon defs set}. Finally, the parameters $\theta^{h*}$ and $\theta^{m*}$ are obtained using a VQE method that minimize the energy of the state prepared by the circuit with respect to the Hamiltonian in Eq.~\eqref{eq: Z2 Ham MGF JW in Hh Hm He}.}
    \label{fig: GS circuit}
\end{figure}
In this section, we use the Hamiltonian and the ansatz for hadron states defined in Sec.~\ref{sec: Theoretical formalism} to lay out a digital quantum algorithm for performing multi-hadron scattering. 
We choose to work with the MGF for a more efficient circuit implementation.

The overall protocol used in this work starts from the SCV state $\ket{\Omega}_\text{SCV}$ in Eq.~\eqref{eq: SCV in MGF def}, which is one of the computational basis states.
The protocol is composed of three circuit modules: 
\begin{enumerate}
    \item $\mathcal{Q}_{\rm GS}$ prepares the interacting vacuum $\ket{\Omega}$ from $\ket{\Omega}_\text{SCV}$. 
    \item $\mathcal{Q}_{\rm Init}$ constructs the initial scattering state comprised of well-separated wave packets.
    \item $\mathcal{Q}_{\rm Trott}$ performs the Trotterized time evolution under the Hamiltonian in Eq.~\eqref{eq: Z2 Ham MGF JW in Hh Hm He}.
\end{enumerate}
These three modules are depicted in Fig.~\ref{fig: circuit flow}, with each module separated by a dotted line, indicating the corresponding quantum state prepared at the end of each stage.
Each circuit module will be discussed in the following. Computing observables may require its own circuit module, which we will partly discuss in Sec.~\ref{subsec: return probability} and Appendix~\ref{app: Hadamard test}.

Before proceeding, we emphasize that our procedure constitutes a hybrid classical-quantum algorithm, where both the ground state in module 1 and the $k$-momentum eigenstates required in module 2 are prepared using the VQE method.
In this work, the VQE results for the ground state and $k$-momentum eigenstates are verified only through classical evaluation, in order to save quantum-computing resources.
However, in theory, it is possible to implement both VQE calculations on a quantum computer provided that the expectation value of the Hamiltonian operator can be determined with sufficient accuracy from the hardware results (which generally require a large number of measurements).

\subsection{$\mathcal{Q}_{\rm GS}$: Preparing the interacting vacuum
\label{subsec: QGS for interacting vacuum}}

The circuit for $\mathcal{Q}_{\rm GS}$ is similar to the one given in Ref.~\cite{Davoudi:2024wyv} (which was inspired by a VQE ground-state circuit in the case of a (2+1)D $Z_2$ LGT in Ref.~\cite{Lumia:2021tpu}). The strategy is to prepare the ground state via iterative evolution of an initial state using the terms in the Hamiltonian.
Thus, $\mathcal{Q}_{\rm GS}$ is parameterized by $\theta^h_x$, $\theta^m_x$, and $\theta^\epsilon_x$, which are related to $H^h$, $H^m$, and $H^\epsilon$ in Eqs.~\eqref{eq: Hh def}-~\eqref{eq: Hepsilon def}, respectively.
The form of $\mathcal{Q}_{\rm GS}$ is given by
\begin{equation}
    \mathcal{Q}_{\text{GS}} = \prod_{x=1}^{N_{\rm GS}} \left(\prod_{n \in \Gamma}{e^{-\frac{i}{2}\theta^{h}_x H_{n,n+1}^{h} }}\right) \left(\prod_{n \in \Gamma}e^{-\frac{i}{2}\theta^{m}_x H^{m}_n }\right) \left(\prod_{n \in \Gamma }e^{-\frac{i}{2}\theta^{\epsilon}_x H^{\epsilon}_n }\right),
    \label{eq: QGS def}
\end{equation}
where $N_{\rm GS}$ denotes the number of iterations and
\begin{subequations}
\begin{align}
    &H^h_{n,n+1} = \begin{cases}
        \sigma^{\textbf{x}}_{n}\sigma^{\textbf{x}}_{n+1} + \sigma^{\textbf{y}}_{n}\sigma^{\textbf{y}}_{n+1} \quad \text{if} \quad n \neq N-1, \vspace{10pt} \\
        \alpha_N\left({ \sigma^{\textbf{x}}_{N-1}\tilde{\sigma}^{\textbf{x}}_{N-1}\sigma^{\textbf{x}}_{0}+ \sigma^{\textbf{y}}_{N-1}\tilde{\sigma}^{\textbf{x}}_{N-1}\sigma^{\textbf{y}}_{0}}\right)  \quad  \text{if} \quad n = N-1,
        \end{cases}
    \label{eq: Hh nnp1 def}\\[10pt]
    & H^m_{n} = (-1)^{n+1} \sigma^{\textbf{z}}_{n},
    \label{eq: Hm n def}\\[10pt]
    & H^\epsilon_{n} = \begin{cases}
        \gamma_n\tilde{\sigma}^{\textbf{z}}_{N-1}\left(\prod_{j=0}^n \sigma^{\textbf{z}}_{j}\right) \quad \text{if} \quad n \neq N-1, \vspace{10pt} \\
        \tilde{\sigma}^{\textbf{z}}_{N-1}  \quad  \text{if} \quad n = N-1.
        \end{cases}
    \label{eq: Hepsilon n def}
\end{align}
\end{subequations}
\begin{figure}[t]
    \centering
    \includegraphics[]{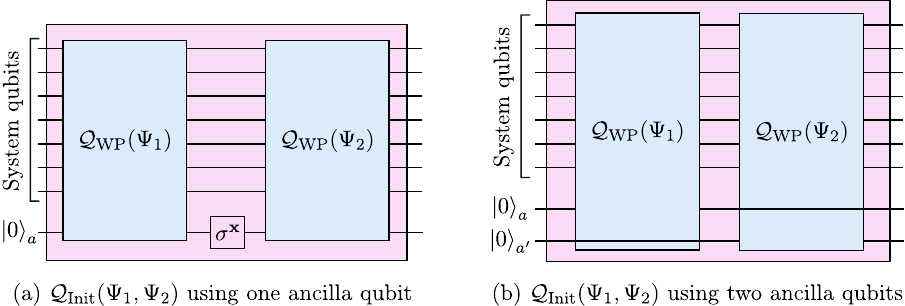}
    \caption{Two different ways of creating the two wave-packet initial scattering state are shown here in (a) and (b). In the former case, the $\mathcal{Q}_{\rm WP}$ circuit that prepares a single-particle wave packet is acted twice using the same ancilla $a_1$ with different wave-packet profiles, $\Psi_1$ and $\Psi_2$, as input. In the latter case, the two applications of $\mathcal{Q}_{\rm WP}$ circuit use two different ancilla $a_1$ and $a_2$. The non-participating ancilla in a $\mathcal{Q}_{\rm WP}$ is indicated with a line crossing over the corresponding circuit block in (b). The method in (a) yields a larger systematic error compared to (b), leading to a lower-fidelity wave-packet state, as discussed in the text. Detailed decomposition of each $\mathcal{Q}_{\rm WP}(\Psi)$ is presented in Appendix~\ref{app: QWP MGF circuit}.
    }
    \label{fig: Q Init circuit}
\end{figure}

It is found that it suffices to set $N_{\rm GS}=1$ and $\theta^\epsilon_1 = 0$ to prepare high fidelity $\ket{\Omega}$ for the parameters $m_f$ and $\epsilon$, and the lattice sizes considered later in Sec.~\ref{sec: Results}. The circuit for $\mathcal{Q}_{\rm GS}$ is shown in Fig.~\ref{fig: GS circuit}. It is parameterized by $\theta^h$ and $\theta^m$, dropping subscript $x=1$ for brevity. In $\mathcal{Q}_{\rm GS}(\theta^h, \theta^m)$, we use the first-order Trotter product formula for each term in $e^{-\frac{i}{2} \theta^h H^h_{n,n+1}}$. Explicitly, we first implement $e^{-\frac{i}{2} \theta^h H^h_{2k,2k+1}}$ for $k=0,\cdots,N/2-1$ before implementing $e^{-\frac{i}{2} \theta^h H^h_{2k+1,2k+2}}$ for $k=0,\cdots,N/2-1$ (with the identification $H^h_{N-1,N} \equiv H^h_{N-1,0}$). This scheme  preserves the global-$Q$ symmetry.
Each Trotter layer requires $8N_P+4$ CNOT gates.
The parameters $\theta^{h*}$ and $\theta^{m *}$ that prepare the ground state are obtained using a VQE algorithm. The algorithm minimizes the energy of the state prepared by $\mathcal{Q}_{\rm GS}(\theta^h, \theta^m)$ with respect to the Hamiltonian in Eq.~\eqref{eq: Z2 Ham MGF JW in Hh Hm He}.
Computing $\braket{H^h}$ from Eq.~\eqref{eq: Hh Hm Hepsilon defs set} requires independent measurements in the $\sigma^{\rm \textbf{x}}$ and $\sigma^{\rm \textbf{y}}$ basis for qubits encoding fermionic fields, while the qubit encoding the boson field needs to be measured only in the  $\tilde{\sigma}^{\rm \textbf{x}}$ basis. The values of $\braket{H^m}$ and $\braket{H^\epsilon}$ in the state prepared by $\mathcal{Q}_{\rm GS}(\theta^h, \theta^m)$ can be measured simply in the computational basis of the qubits.

\subsection{$\mathcal{Q}_{\rm Init}$: Preparing the initial scattering state
\label{subsec: QInit for initial state}}

The second module, $\mathcal{Q}_{\rm Init}(\Psi_1, \Psi_2)$, prepares the initial scattering state of two wave packets, $\ket{\Psi_1,\Psi_2}$.
Each wave packet is defined in Eq.~\eqref{eq: WP def} in terms of non-unitary operators $b^\dagger_k$s.
The wave packets thus need to be created by implementing the $b^\dagger_k$ operators in an extended Hilbert space using one or two ancilla qubits, as shown in Fig.~\ref{fig: Q Init circuit}(a) and (b), respectively.
Here, $\mathcal{Q_{\rm WP}}(\Psi_i)$ creates a single wave packet state $\ket{\Psi_i}$ using an ancilla qubit that is initiated in $\ket{0}$. 
The method in Fig.~\ref{fig: Q Init circuit}(a) uses only one ancilla qubit for both $\mathcal{Q_{\rm WP}}(\Psi_1)$ and $\mathcal{Q_{\rm WP}}(\Psi_2)$, with a $\sigma^{\textbf{x}}$ operator acting on the ancilla in between the two circuits. In Fig.~\ref{fig: Q Init circuit}(b), on the other hand, each  $\mathcal{Q_{\rm WP}}$ uses a different ancilla qubit.
Compared to the latter, the former method uses fewer quantum resources but at a cost of lower accuracy of achieving the target state, as will be discussed below.

The circuit for $\mathcal{Q_{\rm WP}}$ is based 
on that in Ref.~\cite{Jordan:2011ci} for the momentum creation operators $b^\dagger_k$ in a non-interacting scalar field theory, in which case $b^\dagger_k$'s exact expressions are known.
The method relies on two properties of $b^\dagger_k$: the operator obeys $\left[b_k,b^\dagger_{k'} \right]=\delta_{k,k'}$, and $b_k\ket{\Omega}=0$.
It is expected that the $b^\dagger_k$ operators in the interacting theory, obtained via the ansatz in Eq.~\eqref{eq: bk dagger complete ansatz definition}, obey these relations only approximately. Then from Eq.~\eqref{eq: b psi dagger def}, $b_\Psi \ket{ \Omega}\approx 0$ and $[b_\Psi, b_\Psi^{\dagger}] \approx \mathds{1}$ for a wavefunction profile $\Psi(k)$ that is normalized as $(\Psi|\Psi)=1$, where 
\begin{align}
(\Psi_2|\Psi_1) \coloneq \sum_{k\in\widetilde{\Gamma}} \Psi^*_2(k)\Psi_1(k).
\label{eq:overlap-def}
\end{align}
With this, one can introduce an ancilla qubit, $a$, and define the following Hermitian operator:
\begin{align}
    \Theta_{\Psi} \coloneq  b_\Psi^{\dagger} \otimes |1_{a} \rangle \langle 0_{a}| +  b_\Psi \otimes |0_{a} \rangle \langle 1_{a}|
    \label{eq: Theta Psi def}
\end{align}
such that
\begin{equation}
    e^{-i \frac{\pi}{2} \Theta_{\Psi}} \ket{\Omega} \otimes \ket{0_{a}} = -i \ket{\Psi} \otimes \ket{1_{a}}.
    \label{eq: ancilla trick}
\end{equation}

The circuit block $\mathcal{Q}_{\rm WP}(\Psi_1)$ in Fig.~\ref{fig: Q Init circuit} implements Eq.~\eqref{eq: ancilla trick} for the wave-packet profile $\Psi_1(k)$. The circuit acts upon qubit $a_1$ initialized in $\ket{0_{a_1}}$, and on the system qubits holding the interacting vacuum $\ket{\Omega}$. Recall that $\ket{\Omega}$ is prepared by $\mathcal{Q}_{GS}(\theta^{h*}, \theta^{m*})$ in the previous step. Furthermore, 
The unitary operator $e^{-i\frac{\pi}{2}\Theta_\Psi}$ is implemented via a Trotter expansion, where the terms constituting  $\Theta_{\Psi}$ are exponentiated separately a number of times with the angle $\pi/2$ divided into a number of Trotter steps $\tilde{n}_t$.
Design and implementation of $\mathcal{Q}_{\rm WP}$ is presented in our earlier work~\cite{Davoudi:2024wyv}, although within the EGF of the $Z_2$ LGT.
The methods used there can still be directly translated to the MGF, as summarized in Appendix~\ref{app: QWP MGF circuit}.
We have restricted our discussion to the case of a $j=1$ order ansatz, and using a second-order product formula with one Trotter step only.
In general, preparing a single wave packet at order $j$ (once the ansatz optimization is done classically or via VQE) requires $[4(j^2+9j+1)N_P+2j^2+2j]\times2\tilde{n}_t$ CNOT gates for an $N$-site theory.

After preparing the first wave packet  $\ket{\Psi_1}$ in the initial scattering state, the remaining part of the second module is achieved by applying the second $\mathcal{Q}_{\rm WP}$ either using the same ancilla, following the action of $\sigma^{\textbf{x}}_{a_1}$, as shown in Fig.~\ref{fig: Q Init circuit}(a), or by using a second ancilla, $a_2$, initiated in $\ket{0_{a_2}}$, as shown in Fig.~\ref{fig: Q Init circuit}(b).
In both cases, the state becomes $-i\ket{\Psi_1}\otimes \ket{0_{a'}}$ with $\ket{0_{a'}}$ being the corresponding ancilla qubit.
The second wave packet $\Psi_2$ can then be prepared by applying $\mathcal{Q}_{WP}(\Psi_2)$ on the state $\ket{\Psi_1}\otimes \ket{0_{a'}}$, provided that the overlap between two wave packets is negligible, i.e., $\left(\Psi_2|\Psi_1\right) \approx 0$, a condition that holds for far separated wave packets.
This statement can be proved by noting that
\begin{align}
   e^{-i\theta\Theta_{\Psi_2}} \ket{\Psi_1}\otimes\ket{0_{a'}} &
   \approx \cos(\theta) \ket{\Psi_1}\otimes \ket{0_{a'}} -i \sin(\theta) \ket{\Psi_1,\Psi_2} \otimes \ket{1_{a'}} \nonumber\\
    &+ (\Psi_2|\Psi_1) \left[\cos(\sqrt{2}\theta) - \cos(\theta) \right] \ket{\Psi_2}\otimes \ket{0_{a'}} \nonumber\\
    &-i  (\Psi_2|\Psi_1) \left[\frac{1}{\sqrt{2}}\sin(\sqrt{2}\theta) - \sin(\theta) \right] \ket{\Psi_2,\Psi_2} \otimes \ket{1_{a'}},
    \label{eq: ancilla trick for two WPs}
\end{align}
where $\ket{\Psi_1,\Psi_2} \coloneq b_{\Psi_2}^\dagger b_{\Psi_1}^\dagger \ket{\Omega}$ and $\ket{\Psi_2,\Psi_2} \coloneq b_{\Psi_2}^\dagger b_{\Psi_2}^\dagger \ket{\Omega}$. The right-hand side of Eq.~\eqref{eq: ancilla trick for two WPs} is derived by expanding the left-hand side, repeatedly using $\left[b_{k_1},b^\dagger_{k_2} \right] \approx \delta_{k_1,k_2}$ and $b_k\ket{\Omega} \approx 0$, and imposing the $\Psi_1$ and $\Psi_2$ normalizations $(\Psi_1|\Psi_1) = (\Psi_2|\Psi_2) = 1$.
Thus, upon taking $\theta = \pi/2$ and assuming $(\Psi_2|\Psi_1) \approx 0$, one arrives at
\begin{equation}
   \mathcal{Q}_{\rm WP} (\Psi_2) \ket{\Psi_1}\otimes\ket{0_{a'}} 
   \approx -i  \ket{\Psi_1,\Psi_2} \otimes \ket{1_{a'}}.
    \label{eq: action of QWP on Psi1}
\end{equation}
The two-wave-packets state is, therefore, expected to be prepared by $Q_{\rm Init}(\Psi_1, \Psi_2)$ if the ancilla qubit(s) is (are) measured in the state $\ket{1_{a_1}}$ ($\ket{1_{a_1}1_{a_2}}$) when prepared using one (two) ancilla(s).
However, the imperfection in ansatz leads to deviation from this expectation, i.e., there will be a non-zero probability of preparing states other than the wave-packet state when the ancilla(s) measurement turn in the outcome $\ket{1_{a_1}}$ ($\ket{1_{a_1}1_{a_2}}$). Furthermore, the hardware implementation of $Q_{\rm Init}(\Psi_1, \Psi_2)$ will incur the device noise, which could contribute to an error compounded with the systematic error.
We refer to this combined error as the ancilla-violation error, and denote it with $\slashed{a}$ in the following.

\begin{figure}[t]
    \centering
    \includegraphics[]{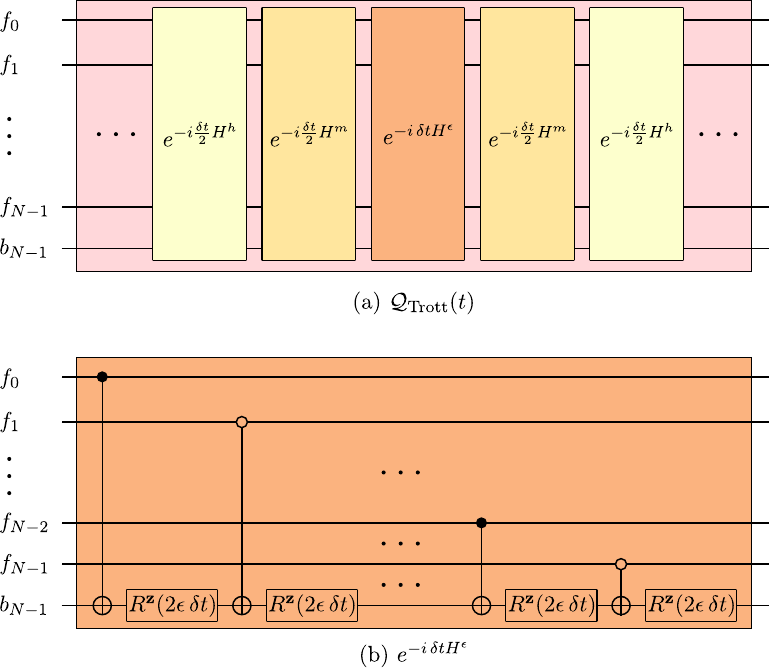}
    \caption{(a) Shown is the Trotterized time-evolution module $\mathcal{Q}_{\rm Trott}$ from Fig.~\ref{fig: circuit flow}, in terms of its constituents according to Eq.~\eqref{eq: Trotter time evolution operator U def}.
    The circuit blocks denote one Trotter step and the dots inside the circuit indicate repeated application of this structure for $n_t$ Trotter step. Circuits for $e^{-\frac{i}{2}H^h \delta t}$ and $e^{-\frac{i}{2}H^m \delta t}$ can be obtained from Fig.~\ref{fig: GS circuit} as discussed in the text, and the circuit for $e^{-i \epsilon H^{\epsilon}\delta t}$ is given in (b). The qubit labels for both (a) and (b), and the single-qubit gate $R^{\textbf{z}}$ have been defined in the caption of Fig.~\ref{fig: GS circuit}. The dotted rows in (b) indicate similar circuit structure for qubits from $f_2$ to $f_{N-3}$. The filled (unfilled) CNOT control indicates that the target $\tilde{\sigma}^{\textbf{x}}$ gate on the boson-link qubit is applied if the control even (odd) matter-site qubit is in $\ket{1}$ ($\ket{0}$) implying a presence of fermion (anti-fermion);  the $R^{\textbf{z}}$ rotation that follows this operation adds the appropriate phase to the state. A detailed explanation of its structure is given in the text.
    \label{fig: H epsilon circuit and Q Trott circuit}}
\end{figure}

Using two ancilla qubits  in $Q_{\rm Init}(\Psi_1, \Psi_2)$ results in a higher-fidelity state than using only one ancilla qubit.
Consider a noiseless scenario, where $\slashed{a}$ denotes purely the systematic error.
In the case of using only $a_1$ for $Q_{\rm Init}(\Psi_1, \Psi_2)$, the probability of ancilla being measured in $\ket{0_{a_1}}$ counts towards $\slashed{a}$.
Thus, if there is an error in the application of the first wave-packet preparation circuit $\mathcal{Q}_{\rm WP}(\Psi_1)$, that is the ancilla-qubit state is a superposition of $\ket{0_{a_1}}$ and $\ket{1_{a_1}}$, this state (upon being hit by $\sigma^{\mathbf{x}}_{a_1}$) propagates through the second wave-packet preparation circuit $\mathcal{Q}_{\rm WP}(\Psi_2)$, and its $\ket{1_{a_1}}$ component contributes to the probability amplitude of $\ket{1_{a_1}}$ in the end of the circuit. This yields a non-zero probability for preparing states other than the two-wave-packet state if ancilla is measured in $\ket{1_{a_1}}$, and such an error, clearly, remains undetected. On the other hand, the $\slashed{a}$ error for $Q_{\rm Init}(\Psi_1, \Psi_2)$ circuit with $a_1$ and $a_2$ is given by the combined probabilities of ancilla qubits being in states $\ket{0_{a_1}0_{a_2}}$, $\ket{0_{a_1}1_{a_2}}$, and $\ket{1_{a_1}0_{a_2}}$.
In this case, such an error can be detected, resulting in higher fidelity for the final state.
We numerically demonstrate these conclusions in Sec.~\ref{subsec: initial state prep results} using a noiseless quantum simulation of $Q_{\rm Init}(\Psi_1, \Psi_2)$ using the two schemes.

\subsection{$\mathcal{Q}_{\rm Trott}$: Trotterized time evolution
\label{subsec: QTrott for time evolution}}

The circuit block $\mathcal{Q}_{\rm Trott}(t)$ in the final module in Fig.~\ref{fig: circuit flow} performs the unitary time evolution, $U(t)\ket{\Psi_1,\Psi_2}$, where $U(t) = e^{-itH}$ with the Hamiltonian $H$ in Eq.~\eqref{eq: Z2 Ham MGF JW in Hh Hm He}.
The part of the circuit shown in red color can be used for computing the return probability, $\mathcal{R}(t) \coloneq |\braket{\Psi_1,\Psi_2|U(t)|\Psi_1,\Psi_2}|^2$, by performing the Hadamard test with an additional ancilla, $a_t$.
It requires a controlled application of the time-evolution operator $U(t)$ with control on the ancilla, as shown in Fig~\ref{fig: circuit flow}.
To study time-dependent expectation values of an operator that is diagonal in the computational basis, one only needs to evolve the system qubits using $U(t)$.
The circuit implementation of $U(t)$ will be discussed here, while the details of the Hadamard test and its extension to controlled operation are left to Appendix~\ref{app: Hadamard test}.

We use the second-order Trotter product formula to approximate $U(t)$, with $\delta t$ as the Trotter time step:
\begin{equation}
    U(t) \approx \prod_{j=1}^{n_t}  \left(e^{-\frac{i}{2}  \delta t H^{h} } \; e^{- \frac{i}{2} \delta t H^{m}} \; e^{-i \delta t H^{\epsilon}} \;e^{- \frac{i}{2}  \delta t H^{m}} \; e^{-\frac{i}{2}  \delta t H^{h}}\right),
    \label{eq: Trotter time evolution operator U def}
\end{equation}
where $t = n_t\,\delta t$, and $H^h$, $H^m$ and $H^\epsilon$ are defined in Eq.~\eqref{eq: Hh Hm Hepsilon defs set}.
Circuits for terms containing $H^m$ and $H^\epsilon$ can be realized without further approximation.
Terms with $H^h$, on the other hand, are implemented by expanding them using the first-order Trotter product formula to separate the terms containing $\sigma^{\textbf{x}}$ operators from the terms containing $\sigma^{\textbf{y}}$ operators.
The circuit blocks for $ e^{-\frac{i}{2}\delta t  H^{h}}$ and $ e^{- \frac{i}{2}\delta t  H^{m}}$ are identical to the subcircuit composed of two-qubit gates and the subcircuit composed of single-qubit gates in $\mathcal{Q}_{\rm GS}$ in Fig.~\ref{fig: GS circuit}, with $\theta^h = \delta t/4$ and $\theta_m = m_f \delta t/2$, respectively.
The circuit for $ e^{-\frac{i}{2} \delta t H^{\epsilon} }$ is shown in Fig.~\ref{fig: H epsilon circuit and Q Trott circuit}, which uses the fact that the phase from $ e^{-\frac{i}{2} \delta t H^{\epsilon}}$ is related to the boson link and the matter distribution across the lattice.
If the fermion occupation is 1 (0) at even (odd) fermion sites, the bosonic qubit is acted with a $\tilde{\sigma}^{\bf x}$ such that the evolved phase is consistent with the Hamiltonian $H^\epsilon$ in Eq.~\eqref{eq: Hepsilon def}.
Furthermore, the conserved charge $Q=N_P$ implies that after the controlled operation on the last fermion-site qubit, the boson-link qubit is restored to its original value, and thus its phase can be evaluated according to the first term in Eq.~\eqref{eq: Hepsilon def}. Each layer of second-order Trotterized time evolution requires $18N_P+8$ CNOT gates.

This concludes the discussion on all the circuit elements involved in preparing the initial scattering state and its time evolution.
In the next section, we present the quantum-emulator and hardware results arising from implementing these circuits to study a scattering process.

\section{Quantum hardware and emulator results
\label{sec: Results}
}

The circuit layout shown in Fig.~\ref{fig: circuit flow} was employed to study scattering in a (1+1)D $Z_2$ LGT with dynamical matter.
The viability of this circuit was demonstrated by executing it on the \texttt{IonQ Forte} quantum computer---a Ytterbium-ion-based quantum computer with 32 qubits, and high-fidelity single- and two-qubit gates~\cite{Chen:2023erd}.
Due to limited device-time availability, the VQE optimizations required for the parameters in the $\mathcal{Q}_{\rm GS}$ circuit, and for the parameters that characterize the ansatz in Eq.~\eqref{eq: bk dagger complete ansatz definition}, as well as the Hadamard test for computing the return probability, were performed using a noiseless quantum-circuit emulator.
The quantum hardware was instead used to implement the vacuum and wave-packet preparation circuits with known parameters, and to evolve the wave packets in a Trotterized manner. A few noise-mitigation strategies are also employed to improve the outcome of the simulations.
All these results are presented and discussed in this section.

We work with two different system sizes, $N_P=5$ and $N_P=13$, corresponding to $N = 10$ and $N=26$, or equivalently $11$ and $27$ qubits, respectively.
The Hilbert space with the symmetry restriction $Q=N_P$ has 504 states for $N_P=5$, and $20,801,200$ states for $N_P=13$.
The former system size is within the reach of the exact-diagonalization technique on a standard computer,
while the low-energy states of the latter system size could be obtained using the DMRG method~\cite{White:1992zz} applied to an MPS ansatz for the states~\cite{Perez-Garcia:2006nqo}, as discussed in Appendix~\ref{app: tensor network}.
We further set $m_f = 1.0$ and $\epsilon = -0.3$, which are the parameters used in our previous work~\cite{Davoudi:2024wyv}. This parameter set ensures that the ratio of the contribution to the vacuum energy from the non-diagonal Hamiltonian, $H^h$, to that from the diagonal Hamiltonian, $H^m+H^\epsilon$, is approximately $0.221$ in the ground state for both $N_P = 5$ and $N_P=13$. This choice, therefore, puts us in a non-trivial regime of parameters, away from the strong- or weak-coupling limits.

\subsection{Variational-quantum-eigensolver optimization
\label{subsec: VQE resutls}
}
The VQE optimization for $\theta^{h*}$ and $\theta^{m*}$ parameters can proceed via the quantum circuit introduced in Sec.~\ref{subsec: QGS for interacting vacuum}. The results for these parameters, as well as the fidelity of the prepared ground state, are summarized in Table~\ref{tab: VQE results GS} in Appendix~\ref{app: VQE tables}.
These results clearly indicate that the ground state prepared after the first module in Fig.~\ref{fig: circuit flow} is a very good approximation of the true ground state $\ket{\Omega}$. Next, the parameters that characterize the ansatz for the low-lying momentum-eigenstate creation operator $b^\dagger_k$ in Eq.~\eqref{eq: bk dagger complete ansatz definition} are obtained for each target momentum $k_t\in \widetilde{\Gamma}$.
One performs a VQE energy minimization for the states arising from $\mathcal{Q}_{\rm WP}(\Psi)$ acted on $\ket{\Omega}$ with $\Psi(k)=\delta_{k,k_t}$ in Eq.~\eqref{eq: WP def}.
We use the second-order Trotter formula on the angle $\pi/2$ in Eq.~\eqref{eq: ancilla trick} with ten and two Trotter steps for $N_P=5$ and $N_P=13$, respectively.
These choices lead to a small Trotter error while keeping the VQE optimization time manageable on the noiseless emulator.
The results of the VQE optimization and the corresponding fidelities are summarized in Tables~\ref{tab: VQE results bk ansatz NP 5} and~\ref{tab: VQE results bk ansatz NP 13} in Appendix~\ref{app: VQE tables}. These results indicate that the target states are captured with a very high fidelity.

\subsection{Preparing the initial hadronic wave packets
\label{subsec: initial state prep results}
}
\begin{table}[t!]
    \renewcommand{\arraystretch}{2.5}
    \begin{center}
        
    \begin{tabular}{|C{1.5cm}|C{1.5cm} C{1.5cm} C{1.5cm}|C{1.5cm} C{1.5cm} C{1.5cm}|C{2cm}|}
    \hline
    $N_P$ & $\sigma_1$ & $\bar{k}_1$ & $\mu_1$ & $\sigma_2$ & $\bar{k}_2$ & $\mu_2$ & $\left(\Psi_2|\Psi_1\right)$\\
    \hline
    \hline
    5 & $\dfrac{7\pi}{20}$ & $\dfrac{2\pi}{5}$ & 2 & $\dfrac{7\pi}{20}$ & $-\dfrac{2\pi}{5}$ & 7 & 0.0666 \\
    \hline
    13 & $\dfrac{3\pi}{13}$ & $\dfrac{2\pi}{13}$ & 6 & $\dfrac{3\pi}{13}$ & $-\dfrac{2\pi}{13}$ & 19 & 0.0104 \\
    \hline
    \end{tabular}
    \end{center}
    \caption{Shown are the parameters of the initial-state wavefunctions defined in Eq.~\eqref{eq: Gaussian wavefunction} for two different lattice sizes. The last column quantifies the overlap of the wavefunctions, with $(\Psi_2|\Psi_1)$ defined in Eq.~\eqref{eq:overlap-def}.
    \label{tab: WP parameters}}
\end{table}
We consider the $i^{\rm th}$ wave packet to have a Gaussian profile
\begin{equation}
    \Psi_i(k)=\mathcal{N}_{\Psi_i} \, \text{exp}\left(-ik\mu_i\right) \text{exp}\left(-\frac{(k-\bar{k}_{i})^2}{4{\sigma_i}^2}\right),
    \label{eq: Gaussian wavefunction}
\end{equation}
with width $\sigma_i$ centered at $\bar{k}_i$ in momentum space, and at $\mu_i$ in position space. $\mu_i$, $\sigma_i$, and $\bar{k}_i$ are all real parameters. The normalization constant $\mathcal{N}_{\psi_i}$ is chosen such that $\left(\Psi_i|\Psi_i\right)=1$.
The two wave packets for the scattering simulation  are centered around opposite momenta with the same magnitude, i.e., $\bar{k}_1=-\bar{k}_2$, such that they move towards each other under time evolution.
Moreover, we set $|\mu_1-\mu_2|=N_P$ such that the wave packets are far separated in position space.
This choice ensures that $\left(\Psi_2|\Psi_1\right)$ is sufficiently small as required for the repeated application of $\mathcal{Q}_{\rm WP}$, see Eq.~\eqref{eq: ancilla trick for two WPs}.
The parameters considered for the $N_P=5$ and $N_P=13$ simulations and the corresponding $\left(\Psi_2|\Psi_1\right)$ values are shown in Table~\ref{tab: WP parameters}. 
\begin{figure}[t]
    \centering
    \includegraphics[scale=1]{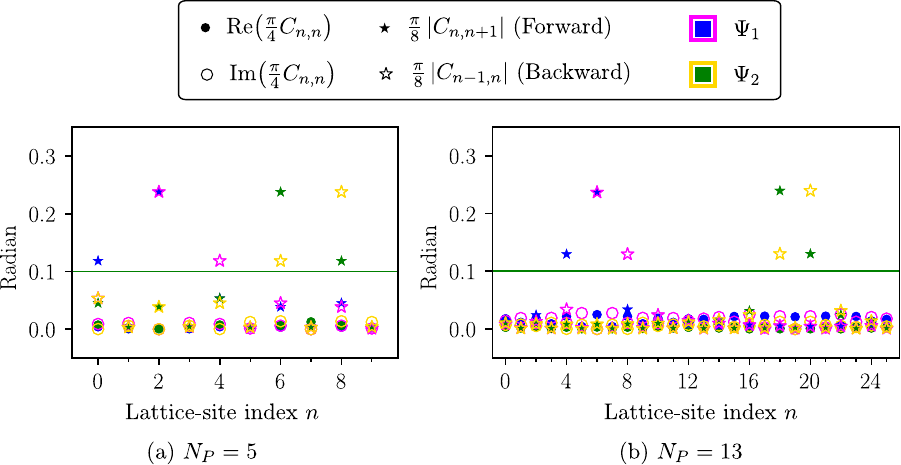}
    \caption{Shown are rotation angles for single-qubit gates appearing in the $\mathcal{Q}_{\rm Init}(\Psi_1,\Psi_2)$ circuit, plotted against their corresponding lattice-site index $n$ for (a) $N_P=5$ and (b) $N_P = 13$. The wave-packet parameters for the Gaussian wavefunctions $\Psi_1(k)$ and $\Psi_2(k)$ are given in Table~\ref{tab: WP parameters}. Together with the optimized ansatz parameters given in Table~\ref{tab: VQE results bk ansatz NP 5} and~\ref{tab: VQE results bk ansatz NP 13}, they determine the coefficients $C_{m,n}$ with $|m-n|\leq1$. The real (imaginary) parts of $C_{n,n}$ coefficients, up to a proportionality constant, are shown in filled (unfilled) circles. Similarly, the magnitude of forward (backward) 1-meson coefficients $C_{n,n+1}$ ($C_{n-1,n}$), up to a proportionality constant, are shown in filled (unfilled) stars.
    The proportionality constants arise from the second-order Trotter expansion used to implement Eq.~\eqref{eq: ancilla trick} with one Trotter step.
    Two different colors are used for each wavefunction to denote the corresponding filled and unfilled markers for better visual differentiability.
    The color used for filled (unfilled) marker is depicted as the color of the inner square (outer border) in the legend.
    The green line shows the cutoff value $\theta_c=0.1$; all 
    rotations angles that fall below this value are discarded when when executing $\mathcal{Q}_{\rm Init}(\Psi_1,\Psi_2)$ on the \texttt{IonQ Forte} quantum computer.
    \label{fig: angle mn plot}
    }
\end{figure}

The optimized parameters from Tables~\ref{tab: VQE results bk ansatz NP 5} and~\ref{tab: VQE results bk ansatz NP 13}, together with the wave-packet-profile parameters in Table~\ref{tab: WP parameters}, can now be used in 
Eq.~\eqref{eq:cmn-def} to compute the $C_{m,n}(\equiv C^{(j)}_{m,n})$ coefficients.
The $j=1$ ansatz considered here contains terms with $m=n$ and 1-length mesons, and the corresponding coefficients determine the rotation angles for the one-qubit gates in $\mathcal{Q}_{\rm Init}(\Psi_1, \Psi_2)$.
We use the second-order Trotter formula with $\tilde{n}_T$ Trotter steps for the angle $\pi/2$ in Eq.~\eqref{eq: ancilla trick}.
Furthermore, to reduce the circuit depth, we choose to implement only the angles $\theta$ such that $\theta>\theta_c$.
The choice of Trotter order, the number of Trotter steps, and $\theta>\theta_c$ impact the accuracy of each wave-packet preparation step.
The implications of such approximations have been thoroughly studied for a single-wave-packet preparation in our previous work~\cite{Davoudi:2024wyv}.
Moreover, the number of ancillary qubits used in $\mathcal{Q}_{\rm Init}(\Psi_1, \Psi_2)$, and the two wave-packets overlap $\left(\Psi_2|\Psi_1\right)$, introduce further systematic errors in the two-wave-packet state preparation, as discussed in Sec.~\ref{subsec: QInit for initial state}.

The various levels of approximations introduce systematic errors in the state preparation. Nonetheless, they offer control over the quantum resources used, i.e., the number of qubits, single-qubit, and two-qubit gates. We consider two different approximations (Appxs) to enable the computations given the quantum resources available to us:
\begin{itemize}
    \item[I.] Using two ancilla qubits, second-order Trotter expansion, $\tilde{n}_T=10$ for $N_P=5$ and $\tilde{n}_T=2$ for $N_P=13$, and $\theta_c=0$,
    \item[II.]  Using one ancilla qubit, second-order Trotter expansion, $\tilde{n}_T=1$ for both system sizes, and $\theta_c=0.1$.
\end{itemize}
The magnitudes of the rotation angles corresponding to Appx II are plotted in Fig.~\ref{fig: angle mn plot}. As is observed from the plotted values, only four $C_{m,n}$ values per wave packet contribute to rotation angles with $\theta > \theta_c=0.1$.\footnote{As is seen from the figure, for the $N_P=13$ system, $\theta_c$ can be set to even smaller values without increasing the number of $C_{m,n}$ coefficients contributing to the $\theta > \theta_c$ set.}

We compare observables in the state $\ket{\Psi_1,\Psi_2}$ prepared with either of these approximations.
Two observables, diagonal in the computational basis, are studied. One is the staggered (fermion) density,   
\begin{align}
    &\chi_n = 
    \begin{cases}
        \braket{\psi_n^\dagger\psi_n} \quad \text{if} \quad n\in \text{even}, \\
        1-\braket{\psi_n^\dagger\psi_n} \quad \text{if} \quad n\in \text{odd},
    \end{cases} \label{eq: staggered density def}
\end{align}
and the other is the electric-field value at the qubit encoding the hardcore boson,
\begin{align}
    E = \braket{\tilde{\sigma}^{\textbf{z}}}
    \label{eq: E field def}.
\end{align}
The quantum resources required to implement $\mathcal{Q}_{\rm Init}(\Psi_1, \Psi_2)$ for the wave-packet parameters listed in Table~\ref{tab: WP parameters} are summarized in Table~\ref{tab: quantum resource init WPs} considering both approximations.
\begin{table}[t!]
    \renewcommand{\arraystretch}{2}
    \begin{center}
    \begin{tabular}{|C{1.5cm}|C{1.5cm}|C{1.5cm} | C{3.2cm}| C{3.2cm}|C{2cm}|}
    \hline
    $N_P$ & Appx & Qubits &  Single-qubit gates (raw/transpiled) & CNOT gates (raw/transpiled) & $\slashed{a}$ \\
    \hline
    \hline
    \multirow{2}{*}{5} & I & 13 & 12757/13343 & 11324/7315 & 13.25\,\%\\
     & II & 12 & 310/369 & 236/167 & 6.96\,\%\\
    \hline
    \multirow{2}{*}{13} & I & 29 & 6925/7660 & 5948/3934 & 11.08\,\%\\
     & II & 28 & 494/572 & 300/197 & 6.07\,\%\\
    \hline
    \end{tabular}
    \end{center}
    \caption{ Shown are the number of qubits, single-qubit gates, and CNOT gates required for preparing the initial scattering state $\ket{\Psi_1,\Psi_2}$ in Fig.~\ref{fig: circuit flow} with the wave-packet parameters in Table~\ref{tab: WP parameters}. The raw gate counts can be obtained from the circuits described in the main text, while the transpiled gate counts are obtained with the \texttt{Qiskit} transpiler. (Version  1.1.1 was used for Appx I while version 1.0.2 was used for Appx II.)
    The two approximations, Appx I and Appx II, yield different  accuracy as described in Sec~\ref{subsec: initial state prep results}.
    Only circuits for Appx II were implemented on the \texttt{IonQ Forte} quantum computer.
    Finally, the last column denotes the ancilla-violation error $\slashed{a}$, calculated using the noiseless \texttt{Aer} simulator (with $5 \times 10^5$ shots). 
    \label{tab: quantum resource init WPs}}
\end{table}

\begin{figure}[t]
    \centering
    \includegraphics[]{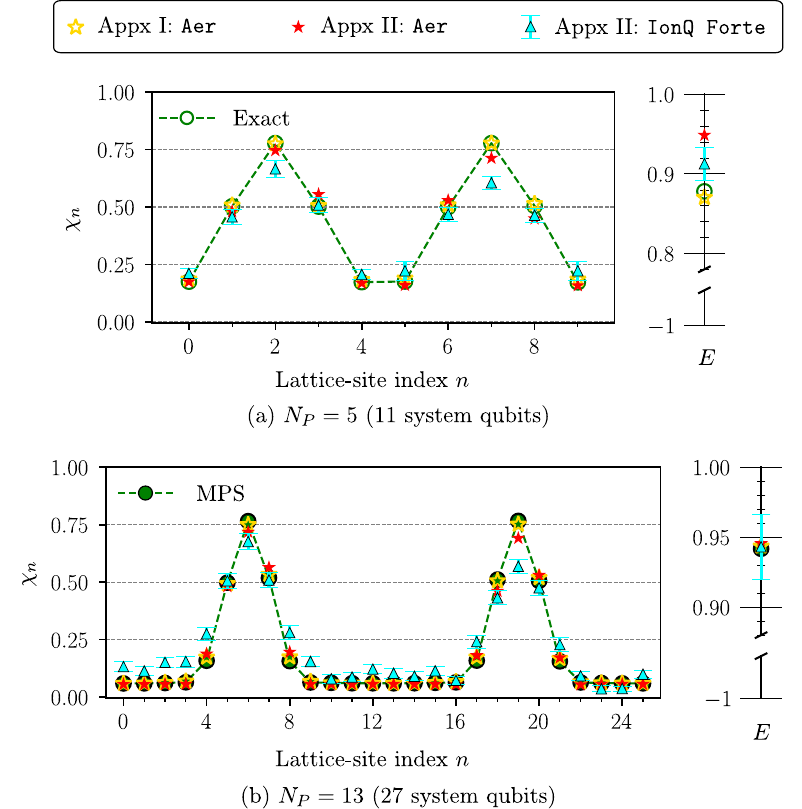}
    \caption{ Shown are the staggered density $\chi_n$ across lattice-site index $n$ (left) and the electric field at the boson qubit $E$ (right) for the two-wave-packet scattering states, with the wave-packet parameters in Table~\ref{tab: WP parameters}. The results for the lattice sizes $N_P = 5$ (top, 11 system qubits) and $N_P = 13$ (bottom, 27 system qubits) are compared against the ideal results, which are obtained by exact numerical solution (empty green circles) and by using the MPS-ansatz states (solid green circles), respectively.
    The dashed green line is depicted for visual guidance and is not a fit to data. Two approximations are implemented: Appx I (yellow stars) and Appx II (red stars), both obtained using \texttt{Aer} noiseless simulator using $5\times 10^5$ shots. Additionally, Appx II is implemented on the \texttt{IonQ Forte} quantum computer (cyan triangles) with 1000 shots for each $N_P$. The error bars 
    are obtained from bootstrap resampling of the global-symmetry-based noise-mitigated hardware results.
    The ranges for the y-axes in all plots are taken to be over all possible values the corresponding observable can take. The axis for $E$ values is broken and rescaled appropriately to resolve the closely located data points near its maximum allowed value. 
    \label{fig: inital state 2WP}
    }
\end{figure}

Results for $\chi_n$ and $E$ are displayed in Fig.~\ref{fig: inital state 2WP}.
These computations are performed using the noiseless \texttt{Aer} simulator using $5\times 10^5$ shots (yielding negligible uncertainty from the shot noise). They are compared against the ideal results obtained from exact numerical results (for $N_P=5$) and using an MPS ansatz for the states (for $N_P=13$).
The Appx I results are in better agreement with the ideal results. Appx II results, nonetheless, are not far off. Adopting Appx II, therefore, is not unreasonable especially since its required resources are significantly lower than Appx I.
These results are obtained by discarding measurements for which the ancilla qubit(s) do not have the correct value(s) $\ket{1_{a_1},1_{a_2}}$ for Appx I and $\ket{1_{a_1}}$ for Appx II.
The percentage of discarded measurements are shown in Table~\ref{tab: quantum resource init WPs} under the column $\slashed{a}$, the ancilla-violation error.
More states are discarded in Appx I than Appx II, since a part of the erroneous states are hidden in the valid ancilla values in Appx II, see discussions in Sec.~\ref{subsec: QInit for initial state}.

The two-wave-packet preparation circuits within Appx II were executed on the \texttt{IonQ Forte} quantum computer, with 1000 shots per circuit. The $\chi_n$ and $E$ values obtained from these runs are shown in Fig.~\ref{fig: inital state 2WP}.
The hardware results contain various errors that are sourced from the trapped-ion quantum devices.
Similar to Ref.~\cite{Davoudi:2024wyv}, we employ a simple post-processing error-mitigation scheme based on the global symmetry of the fermion-qubits configurations. Explicitly, we discard the states from the final results that exhibit $Q\neq N_P$.
Such states indicate the occurrence of at least one error in the fermion qubits. 
We refer to this error as the symmetry-violation error, and denote it by $\slashed{Q}$. We find $\slashed{Q} = 49.40\,\%$ for $N_P=5$ and $\slashed{Q} = 71.20\,\%$ for $N_P=13$.
The states in the set complementary to $\slashed{Q}$ are not necessarily contamination free, since the errors that do not change $Q$ are not filtered away.
Part of these errors percolate into the ancilla-violation error: $\slashed{a} = 14.82\,\%$ for $N_P=5$ and $\slashed{a} = 18.40\,\%$ for $N_P=13$. These are larger than their respective noiseless-simulator results in Table~\ref{tab: quantum resource init WPs}, due to additional hardware errors. The error $\slashed{a}$ is calculated over the shots remained after discarding those with a $\slashed{Q}$ error. States with the incorrect values of ancilla measurement are also discarded before evaluating the observables' expectation-values. 
Finally, the error bars in Fig.~\ref{fig: inital state 2WP} are obtained from the standard deviation of the mean of the bootstrap samples of the physical events with $100$ resampled configurations (at which value the bootstrap-sample mean distributions is stabilized).

The hardware results are in a good agreement with their noiseless-simulator counterparts for both observables.
The deviation from noiseless simulator is generally more prominent at the center of the wave packet. This feature can be attributed to the fact that the qubits at the center of the wave packets are acted on by gates with larger angles (i.e., larger $|C_{m,n}|$ values as seen in Fig.~\ref{fig: angle mn plot}). Larger gate angles indicate longer gate-implementation times, making the gates more susceptible to quantum decoherence.
In summary, the ability to systematically control the various levels of approximation allows us to compromise marginally on the accuracy of preparing the initial scattering state while benefiting from significant reduction in quantum-resource requirements.
This is especially useful if the observable under investigation is local, and is, hence, robust against extensive error accumulation under time evolution~\cite{Trivedi:2022lsi,Kashyap:2024wgf}.
We further investigate this point in Secs.~\ref{subsec: time eovlution of diagonal observables} and~\ref{subsec: return probability} by studying different observables under time evolution.

Finally, our method can be extended to prepare multiple spatially spaced wave packets.
One can repeat the application of $\mathcal{Q}_{\rm WP}(\Psi_i)$ with one or more ancilla qubits as many times as needed. We have demonstrated the preparation of three wave packets in an $N_P=13$ system on the \texttt{IonQ Forte}, see Appendix~\ref{app:three-WP}, observing good agreement with the expected results. 

\subsection{Time-evolved observables
\label{subsec: time eovlution of diagonal observables}
}
\begin{figure}[t]
    \centering
    \includegraphics[]{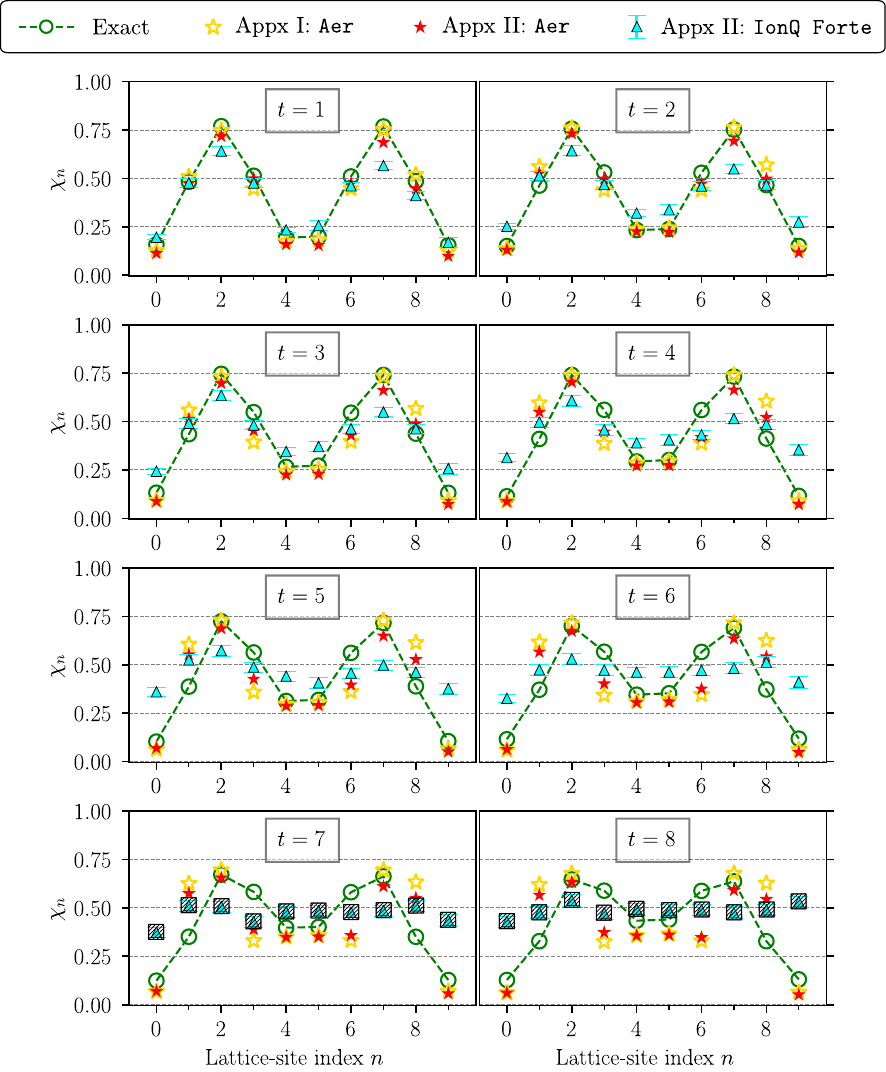}
    \caption{Shown are the expectation values of the staggered density $\chi_n$ across lattice-site index $n$, in a Trotterized time-evolved scattering state of two meson wave packets for $N_P=5$ with 11 system qubits. Each column (row) shares the x-axis (y-axis) label. The plot legends and error bars are the same as in Fig.~\ref{fig: inital state 2WP}. The exact results correspond to the evaluation of time-evolution unitary matrix $U(t) = e^{-itH}$ upon exact exponentiation of the Hamiltonian matrix for a given time $t$, acting on the corresponding initial state in Fig.~\ref{fig: inital state 2WP}(a). The noiseless-simulator and hardware results correspond to $5 \times 10^5$ and $3000$ shots, respectively.
    The quantum circuits for the Trotter time evolution are taken with time steps of $\delta t =1$.
    The meshed squares for the $t=7$ and $t=8$ plots denote the hardware-noise-dominated results. 
    The number of single-qubit and CNOT gates implemented for the combined state-preparation and time-evolution circuits for each $t$ is provided in Table~\ref{tab: device errors}.
    \label{fig: t evolve stagg density}}
\end{figure}
\begin{figure}[t]
    \centering
    \includegraphics[]{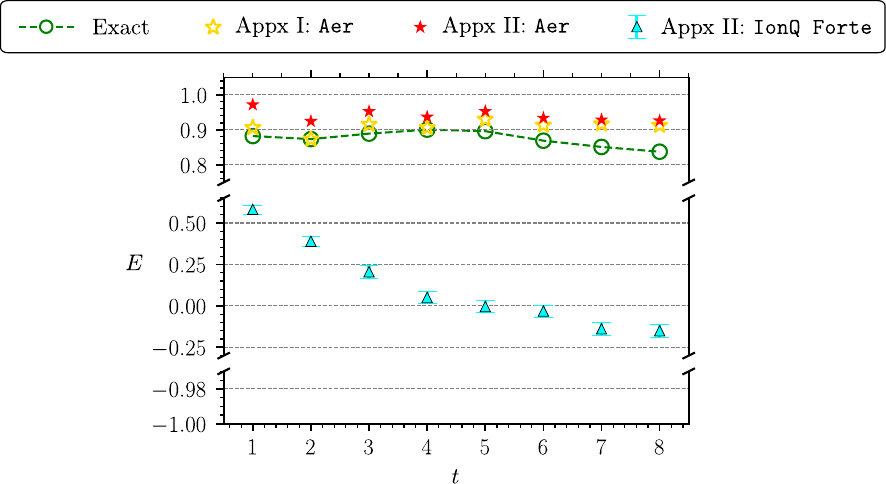}
    \caption{Shown are the expectation values of the electric field $E$ at the boson qubit, in a Trotterized time-evolved scattering state of two wave packets for $N_P=5$ with 11 system qubits.
    The hardware results (cyan triangles) are dominated by noise. 
    Results obtained using the \texttt{IonQ Forte} noisy emulator are shown in Appendix~\ref{app: Pauli twirling and ODR} to demonstrate a possible recovery of the signal using additional noise-mitigation tools, such as Pauli twirling and operator decoherence renormalization, at the cost of additional quantum-processing time.
    \label{fig: t evolve E field}}
\end{figure}
We use the circuit for Trotterized time evolution via a second-order Trotter product formula shown in Sec.~\ref{subsec: QTrott for time evolution} to evolve an initial two-wave-packet state on an \texttt{IonQ Forte} quantum computer.
The $N_P=5$ parameters in Table~\ref{tab: WP parameters} are chosen for this purpose since each Trotter time step involves a shallower circuit than the larger-system counterpart: each Trotter-step circuit block constitutes 204 single-qubit gates and 60 CNOT gates. The three modules in Fig.~\ref{fig: circuit flow} are implemented with Appx II for 3000 shots and $\delta t=1$. The number of shots and the size of the Trotter step required to match the theoretical prediction are estimated using the noisy emulator provided for this device by \texttt{IonQ}.\footnote{For future simulations involving large system sizes, a viable strategy is to repeat experiment for a range of Trotter-step sizes and shot numbers to reach convergence in measured observables.} 

The device results are analyzed using the methods described in the previous subsection, and the $\slashed{Q}$ and $\slashed{a}$ values calculated for each Trotter step are shown in Table~\ref{tab: device errors}.
\begin{table}[t!]
\renewcommand{\arraystretch}{2.2}
\begin{center}
\begin{tabular}{|C{2.9cm}||C{1.35cm}|C{1.35cm}|C{1.35cm}|C{1.35cm}|C{1.35cm}|C{1.35cm}|C{1.35cm}|C{1.35cm}|}
    \hline
    $t$ & 1 & 2 & 3 & 4 & 5 & 6 & 7 & 8 \\ \hline
    Single-qubit gates & 573 & 771 & 969 & 1167 & 1365 & 1563 & 1761 & 1959\\ \hline
    CNOT gates & 227 & 287 & 347 & 407 & 467 & 527 & 587 & 647 \\ \hline
    $\slashed{Q}$ & 56.07\,\% & 62.43\,\% & 67.13\,\% & 70.77\,\% & 71.73\,\% & 73.47\,\% & 73.63\,\% & 76.73\,\% \\ \hline
    $\slashed{a}$ & 15.78\,\% & 17.66\,\% & 16.02\,\% & 19.61\,\% & 18.28\,\% & 19.47\,\% & 20.61\,\% & 20.77\,\% \\ 
    \hline
\end{tabular}
\end{center}
\caption{Shown are the values of the single- and two-qubit gate counts, symmetry-violation error $\slashed{Q}$, and the ancilla-violation error $\slashed{a}$, for different Trotter steps $t$ in time evolution of two meson wave packets, resulted from the \texttt{IonQ Forte} device. The gate counts are associated with the transpiled quantum circuit by the \texttt{Qiskit} transpiler. (The raw gate counts can be obtained from those provided in Table~\ref{tab: quantum resource init WPs} for the state preparation and the cost of each Trotter step of evolution given in the text.)}
\label{tab: device errors}
\end{table}
The growing $\slashed{Q}$ and $\slashed{a}$ errors with each Trotter step indicate increasing noise in the device results with deeper circuits.
As mentioned before, the $\slashed{a}$ error in the noiseless simulation denotes the systematic error due to the approximate nature of $b^\dagger_k$ operators.
In this case, this error does not change with time evolution as the ancilla qubit used for the initial-state preparation does not participate in the time-evolution circuit.
Thus, the increasing $\slashed{a}$ error with each Trotter step in Table~\ref{tab: device errors} purely reflects the increasing hardware error.

The results for the time-evolved staggered density are shown in the Fig.~\ref{fig: t evolve stagg density}.
The values obtained from the quantum circuits are compared against the exact calculations that do not exhibit the Trotter error (i.e., they are obtained by calculating $U(t)$ using exact matrix exponentiation).
From these results, the wave packets can be seen to be moving towards each other with their peak values decreasing with time.
The noiseless results (using $5\times 10^5$ shots) in both approximations (Appxs I and II) follow the exact wave-packet profile in the early times, but deviate from the exact values with increasing evolution time due to accumulated Trotter error. In fact, in some instances, the cruder, hence less resource-intensive, Appx II displays less Trotter error than Appx II. This observation makes it clear that small deviations in the initial states may be insignificant in subsequent simulation steps given the effect of other errors during the evolution.
Nonetheless, the shape of the profile for both initial-state approximations resembles the exact wave-packet evolution, and still offers qualitative description of the physical process.

The hardware results also show the qualitative features of time-evolved staggered density, however, the device noise starts to dominate after Trotter time $t=6$.
Furthermore, our results are obtained without performing any additional error-mitigation-circuit runs and only by discarding the $\slashed{a}$ and $\slashed{Q}$ errors during post-processing. 
Nonetheless, we have checked that discarding the $\slashed{a}$ and $\slashed{Q}$ data does not significantly improve the outcome, and only leads to reduced statistics, hence larger shot noise. The reason can be attributed to the fact that most of symmetry-violated errors are associated with a few (mostly one) bit-flip errors in primarily random locations in the qubit register, whose effect becomes insignificant when computing expectation value of local operators. Our observation is consistent with the conclusions of Ref.~\cite{Nguyen:2021hyk}, which finds that the effect of symmetry-based noise mitigation is both quantity dependent and time dependent.

We further compute the time evolution of the electric-field expectation value at the boson qubit, $E$, as shown in Fig.~\ref{fig: t evolve E field}.
Here, the noiseless result agrees with the exact result up to Trotter errors, and almost retains its value at $t=0$ throughout the evolution. The reason is that flipping the electric field at the boson qubit costs an energy proportional to the system size, see Eq.~\eqref{eq: Hepsilon def}. The hardware result for this quantity, on the other hand, significantly deviates from the exact value. 
Thus, obtaining the Trotter time evolution of $E$ requires further noise-mitigation techniques.
We have demonstrated this in Appendix~\ref{app: Pauli twirling and ODR} using the \texttt{IonQ Forte} noisy emulator where Pauli twirling~\cite{Wallman:2015uzh}, along with operator decoherence renormalization~\cite{Urbanek:2021oej,ARahman:2022tkr,Farrell:2023fgd}, are employed to recover the time-evolved value of $E$.\footnote{We relied on the noisy emulator for this analysis because such mitigation techniques require more quantum processing, and our access to the device was limited.}

The hardware runs' moderate coherence time, and time evolution's large circuit depth, prohibit accessing interesting long-time scattering dynamics in this hardware study. Another issue is the small system size, leading to boundary affects in the simulation outcome in the long-time limit. One would, therefore, need to simulate evolution of wave packets in larger systems, but the associated circuit depths would increase considerably. Nonetheless, the results presented here marks the first hadron-scattering simulation on a quantum computer;\footnote{See also the parallel submission by Schuhmacher et al. in the same \texttt{arXiv} listing.} it has pushed the limits of what is possible for such an involved simulation problem on any quantum hardware to date.

\subsection{Return probability
\label{subsec: return probability}
}
\begin{figure}[t]
    \centering
    \includegraphics[]{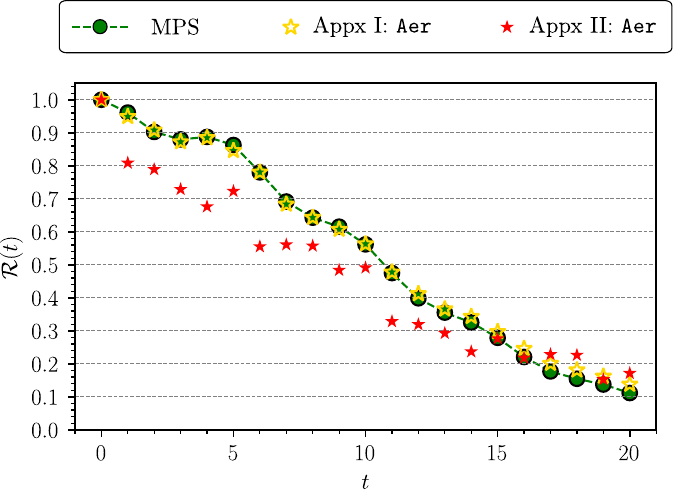}
    \caption{Shown is the return probability, $\mathcal{R}(t)$, of the initial two wave-packet state for $N_P=13$ (27 system qubits) against time. The Trotter time step is $\delta t=0.25$, but results for only integer times are plotted. The MPS results are obtained using the TDVP algorithm, and the values for Appxs I and II cases are calculated using the circuit for the Hadamard test. 
    \label{fig: t evolve return probability}}
\end{figure}
Computing the scattering $S$-matrix, which relates scattering states in early and late times, is a critical observable in nuclear and high-energy physics.
In this section, we compute the return probability (also known as the survival probability and Loschmidt echo):
\begin{equation}
    \mathcal{R}(t) \coloneq \left|\braket{\Psi_1,\Psi_2|U(t)|\Psi_1,\Psi_2}\right|^2,
    \label{eq: return probability def}
\end{equation}
which is a diagonal entry of the scattering $S$-matrix. 
Here, we restrict our discussion to the computation of return probability for the initial state consisting of the two wave packets prepared in Sec.~\ref{subsec: initial state prep results}, rather than its phenomenological implications, or more interesting, but significantly more involved, final-state-momentum-dependent overlaps.

Consider the case of $N_P=13$ with the initial state $\ket{\Psi_1,\Psi_2}$ shown in Fig.~\ref{fig: inital state 2WP}(b), using both approximations, Appxs I and II.
Recall that the former (latter) approximation requires more (less) quantum resources, and prepares a state with higher (lower) fidelity with respect to the target state. 
The return probability for each approximation is computed using the Hadamard test, shown schematically in Fig.~\ref{fig: circuit flow} and described in more detail in Appendix~\ref{app: Hadamard test}, using the noiseless \texttt{Aer} emulator.
The Trotter step size is taken to be $\delta t = 0.25$ to avoid a large Trotter error, and the results are shown in Fig.~\ref{fig: t evolve return probability} only for integer times to avoid clutter.
The emulator results are compared against the ideal results obtained from optimizing an MPS ansatz for the state and performing the time evolution using the time-dependent variational-principle (TDVP) algorithm~\cite{Haegeman:2011zz,Haegeman:2016gfj,Yang_2020}.

Figure~\ref{fig: t evolve return probability} indicates that the return probability associated with Appx II deviates significantly from the ideal case, but that associated with Appx I agrees with the ideal result up to Trotter errors.
This can attributed to the non-local nature of the observable $\mathcal{R}(t)$ which is sensitive to the state vector.
Large deviations from the target state vector, like in Appx II, cause interference effects from the contamination when computing the probability in Eq.~\eqref{eq: return probability def}.
This is in contrast to local observables that are rather insensitive to small deviations from the target state, see Appendix~\ref{app: NP13 local observables} for the example of local staggered density for the same simulation.

Preparing a high-fidelity initial state is, therefore, of crucial importance in future precision-physics applications of quantum computers, especially for observables like the $S$-matrix elements that are highly sensitive to the state vector.
On the other hand, understanding qualitative behavior of local observables under time evolution is a more tractable problem for the NISQ devices, as demonstrated in Sec.~\ref{subsec: time eovlution of diagonal observables}.

\section{Conclusion and outlook
\label{sec: Conclusion and Outlook}}
We developed and implemented a digital algorithm for simulating scattering of hadrons on a quantum computer.
Focusing on a (1+1)D $Z_2$ lattice gauge theory with fermionic matter, we first prepared the initial scattering state of well-separated single-hadron wave packets with well-defined localized momenta.
Our hybrid classical-quantum state-preparation  algorithm offers control over various levels of approximations for preparing the initial state, such that the fidelity with the target state can be systematically improved at the cost of increasing quantum resources.
We prepared the two- and three-wave-packet state on the \texttt{IonQ Forte} trapped-ion-based quantum computer for systems with $10$ and $26$ fermionic sites, using one of these approximations, requiring significantly lower circuit depths, see Table~\ref{tab: quantum resource init WPs}. 
Even in the presence of hardware noise, results agree well with the ideal results, see Figs.~\ref{fig: inital state 2WP} and~\ref{fig: 3WP}, and a symmetry-based noise mitigation yielded only small improvement.

We then performed a quantum time evolution using the second-order Trotter product formula for a system of $10$ fermionic sites, which allowed for hadrons to evolve till they collide, see Fig.~\ref{fig: t evolve stagg density}.
Unfortunately, the device noise washed out the signal for the local fermion-density observable before the hadronic wave packets could fully collide.
Nonetheless, the early-evolution time data are encouraging, despite a large number of entangling gates used and no major noise-mitigation strategy implemented.
The electric-field energy stored in the remnant gauge link (after gauge fixing) exhibited significant decoherence even at the early steps of time evolution, see Fig.~\ref{fig: t evolve E field}.
Nonetheless, using a noisy emulator, we demonstrated the effectiveness of two noise-mitigation schemes, Pauli twirling together with operator decoherence renormalization, in recovering the electric-field value.

Finally, we emphasized the importance of preparing high-fidelity initial scattering states for future precision computations of state-sensitive observables, such as phase shifts, $S$-matrix, and decay width.
Our computation of the return probability, which is a diagonal entry of the $S$-matrix, demonstrates that approximations that may appear benign when measuring local observables, appear to exacerbate the error in overlap quantities such as return probability, see Fig.~\ref{fig: t evolve return probability}.

In the following, we present an outlook of this study, including potential avenues for future research:
\begin{itemize}
    \item[$\diamond$] Our ansatz is both effective and efficient in creating single-particle hadronic states with high fidelity in (1+1)D Abelian theories (see our previous work for a U(1) example~\cite{Davoudi:2024wyv}).
    Given the ultimate goal of preparing hadronic states in QCD on a quantum computer, the natural next step is to extend the ansatz to non-Abelian LGTs, and to LGTs in more than one spatial dimensions. These involve introducing additional gauge-invariant bare-meson operators, both arising from non-Abelian components of the fields, and from Wilson-loop operators in (2+1) and higher dimensions. One hopes that the ansatz can approximate the target hadronic state with high fidelity using only a polynomial number of parameters, even toward the continuum limit.
    
    \item[$\diamond$] Quantum simulation of non-Abelian LGT dynamics in more than one spatial dimensions is a resource-intensive task~\cite{Byrnes:2005qx,Shaw:2020udc, Ciavarella:2021nmj, Kan:2021xfc,Lamm:2019bik,Haase:2020kaj,Davoudi:2022xmb,Murairi:2022zdg,Rhodes:2024zbr,Lamm:2024jnl,Balaji:2025afl}. 
    It is perceivable that various alternative formulations of the Kogut-Susskind Hamiltonian formulation of LGTs~\cite{Banuls:2017ena, Raychowdhury:2019iki, Kadam:2022ipf,Kadam:2024zkj, Pardo:2022hrp, Davoudi:2020yln, Chandrasekharan:1996ih, Wiese:2021djl, Zohar:2014qma,Zohar:2018cwb,Zohar:2019ygc,Alexandru:2023qzd,Alexandru:2019nsa,Ji:2022qvr,Ji:2020kjk,Kavaki:2024ijd,Illa:2025dou,DAndrea:2023qnr,Zache:2023dko, Hartung:2022hoz,Romiti:2023hbd, Fontana:2024rux, Bauer:2021gek,Grabowska:2024emw,Burbano:2024uvn,Ciavarella:2025bsg,Ciavarella:2021nmj} will alleviate some of the resource requirements.
    Among these, the loop-string-hadron (LSH) formulation~\cite{Raychowdhury:2019iki,Kadam:2022ipf, Kadam:2024zkj}, is particularly appealing since its degrees of freedom are generated by fully local and gauge-invariant operators in the electric basis, and only an algebraic Abelian constraint remains to be enforced on each link of the lattice. This formulation, therefore, may present a cost-efficient framework for extending our ansatz. Bare-meson operators can be simply mapped to gauge-invariant states of the LSH formulation, and the Abelian constraints can aid noise mitigation in quantum-circuit implementations. 

    \item[$\diamond$] The single-hadron momentum eigenstates prepared using the algorithm of this work, and extended to more complex gauge theories, are crucially needed in studying hadron structure using quantum computers. Ultimately, one aims to constrain, from real-time correlators of hadrons, various non-perturbative QCD matrix elements of relevance to collider physics~\cite{Bauer:2025nzf}, such as various distribution functions~\cite{Lamm:2019uyc,Echevarria:2020wct,Li:2021kcs,Mueller:2019qqj,Perez-Salinas:2020nem,Qian:2021jxp,Pedernales_2014,Gustin:2022pfu} for large-hadron and electron-ion colliders. The expressive form of our ansatz, and its potential extensions, may prove valuable in gaining more insights into such structure quantities in the future.
    
    \item[$\diamond$] Our method is applicable to preparation of any number of hadronic wave-packets (as long as systems size is large enough to contain well-separated wave packets). 
    This feature is particularly useful when one aims to compute exclusive scattering amplitudes. A second qubit register is used to prepare the desired multi-hadron final state. Then a swap test~\cite{Buhrman:2001rma} can be used to compute the overlap between the time-evolved initial state and the desired final state. 
    Upon insertion of spacetime-separated electroweak currents, one can further access quantities such as hadron tensor. Thus, the optical theorem can be invoked to obtain the inclusive scattering cross section for a given center-of-mass energy.
    However, both of these analyses require larger lattice sizes and evolution times, hence wider and deeper quantum circuits.

    \item[$\diamond$] The scattering protocol presented here allows one to access the outgoing scattering state at any moment past collision. Therefore, not only can one access asymptotic observables such as the $S$-matrix, but also can measure other local and non-local observables in early moments after the collision and beyond. A novel quantity of interest, not readily accessible in collider experiments, is entanglement, in form of bipartite entanglement entropy or entanglement Hamiltonian. There is significant interest in quantum simulations of string breaking, hadronization, fragmentation, and thermalization in the aftermath of high-energy hadronic collisions~\cite{Cochran:2024rwe,De:2024smi,Gonzalez-Cuadra:2024xul,Ciavarella:2024lsp,Crippa:2024hso,Liu:2024lut,Alexandrou:2025vaj,Luo:2025qlg,Zhou:2021kdl,Mueller:2024mmk,Li:2024nod,Farrell:2024mgu,Lee:2024jnt,Florio:2024aix,Davoudi:2024osg}. Nonetheless, many of early experimental studies are limited to simple models and scenarios, and do not involve high-energy hadronic collisions. Future quantum-simulation studies, built upon efficient high-energy hadronic-state preparation schemes (such as that presented in this work), can begin to shed light on such important non-equilibrium phenomenology.
\end{itemize}

\noindent
\emph{Note:} A parallel submission by J.~Schuhmacher, G.~X.~Su, J.~J.~Osborne, A.~Gandon, J.~C.~Halimeh, and I.~Tavernelli is to appear in the same \texttt{arXiv} listing as our paper, where authors perform a quantum-simulation experiment of hadron scattering in a quantum-link-model formulation of the U(1) LGT on an \texttt{IBM} quantum computer.

\section*{Acknowledgment}
Quantum-emulator and quantum-hardware computations of this work were enabled by access to IonQ systems provided by the University of Maryland's National Quantum Laboratory (QLab). We thank support from IonQ scientists and engineers, especially Daiwei Zhu, during the execution of our runs. We further thank Franz Klein at QLab for the help in setting up access and communications with IonQ contacts.
S.K. acknowledges valuable discussions with Roland Farrell and Francesco Turro.
This work was enabled, in part, by the use of advanced computational, storage and networking infrastructure provided by the Hyak supercomputer system at the University of Washington.
The numerical results in this work were obtained using \texttt{Qiskit}~\cite{Qiskit}, \texttt{ITensors}~\cite{ITensor}, \texttt{Julia}~\cite{Julia-2017} and \texttt{Jupyter Notebook}~\cite{kluyver2016jupyter} software applications within the \texttt{Conda}~\cite{conda} environment.

Z.D. and C-C.H. acknowledge support from the National Science Foundation's Quantum Leap Challenge Institute on Robust Quantum Simulation (award no. OMA-2120757); the U.S. Department of Energy (DOE), Office of Science, Early Career Award (award no. DESC0020271); and the Department of Physics, Maryland Center for Fundamental Physics, and College of Computer, Mathematical, and Natural Sciences at the University of Maryland, College Park. Z.D. is further grateful for the hospitality of Nora Brambilla, and of the Excellence Cluster ORIGINS at the Technical University of Munich, where part of this work was carried out.
The research at ORIGINS is supported by the Deutsche Forschungsgemeinschaft (DFG, German Research Foundation)
under Germany's Excellence Strategy (EXC-2094–-390783311).
S.K. acknowledges support by the U.S. DOE, Office of Science, Office of Nuclear Physics, InQubator for Quantum Simulation (IQuS) (award no. DE-SC0020970), and by the DOE QuantISED program through the theory consortium ``Intersections of QIS and Theoretical Particle Physics'' at Fermilab (Fermilab subcontract no. 666484).
This work was also supported, in part, through the Department of Physics and the College of Arts and Sciences at the University of Washington.

\appendix

\section{Verifying the ansatz validity for a larger system using tensor networks
\label{app: tensor network}
}

In this appendix, we study the applicability of our wave-packet ansatz to a larger system than considered in Sec.~\ref{subsec: ansatz continuum limit numerical analysis}. 
We consider the $Z_2$ LGT on a lattice with 26 staggered sites (13 physical sites) with the Brillouin zone $\tilde{\Gamma} = \{ -\frac{6\pi}{13},-\frac{5\pi}{13}, \cdots, \frac{6\pi}{13} \}$. 
The true momentum eigenstates, while cannot be obtained by exact diagonalization due to the size of the Hilbert space, can be approximated with a matrix product states (MPS) ansatz on a tensor network. 
Using the density-matrix-renormalization-group (DMRG) optimization, one can obtain $\ket{\Omega}_{\rm MPS}$ that recovers the exact interacting ground state $\ket{\Omega}$ with high precision in (1+1)D. 
Excited states can also be built successively using the same DMRG process upon constraining the output to be orthogonal to the ground state and all other excited states with lower energy, if any. 
The MPS states are further constrained such that the total fermionic excitation remains in the $Q=N_P$ sector, and thus the gauge symmetry is guaranteed on a periodic lattice. 
All DMRG calculations in this work are carried out using the \texttt{ITensors} library~\cite{ITensor}.

The DMRG calculations for the ground state and the 13 excited states with the lowest energy in each momentum sector are performed for a range of $(m_f, \epsilon)$ values. 
In each DMRG optimization, the maximum allowed bond dimension is set to 600, and the Schmidt coefficients under $10^{-12}$ are truncated. 
The number of sweeps is set to $\mathcal{O}(100)$ to ensure convergence in energy. 
To further verify the quality of an output DMRG state $\ket{\psi}_{\rm MPS}$, the variance of energy is calculated:
\begin{equation}
    {\rm Var}
    \coloneq \frac{{}_{\rm MPS}\langle \psi{} |H^2|\psi\rangle_{\rm MPS}- \left( {}_{\rm MPS}\langle \psi |H|\psi\rangle_{\rm MPS} \right)^2}{\left( {}_{\rm MPS}\langle \psi |H|\psi\rangle_{\rm MPS} \right)^2}.
\end{equation}
If $\ket{\psi}_{\rm MPS}$ is exactly an eigenstate of $H$, the variance should be zero. 
For all pairs $(m_f, \epsilon)$, the DMRG parameters listed above can achieve ${\rm Var}=\mathcal{O}(10^{-10})$.
\begin{figure}[t!]
    \centering
    \includegraphics[scale = 0.9]{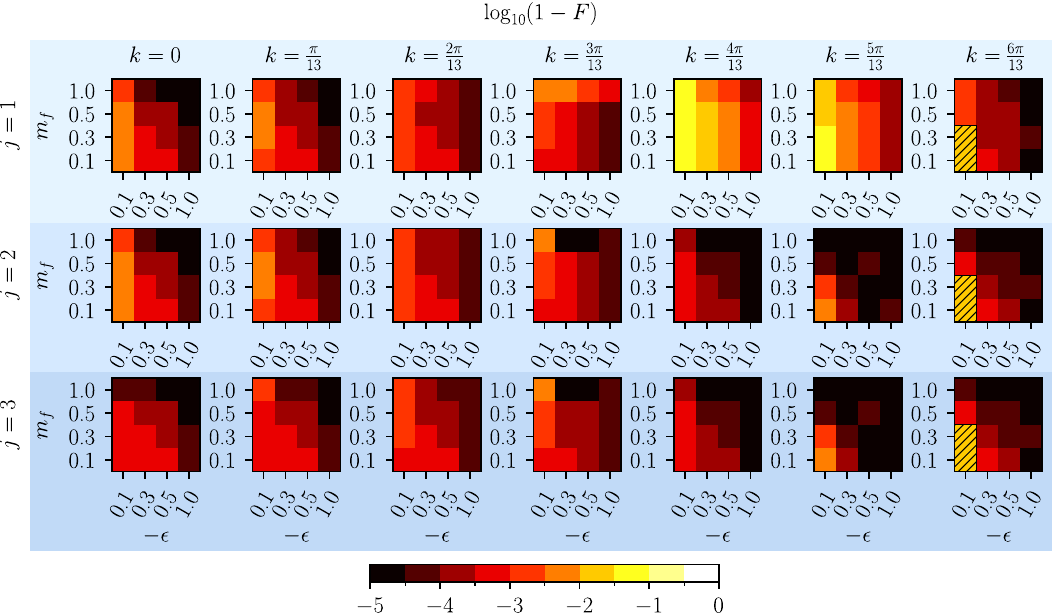}
    \caption{The parameter scan for the 26-site (1+1)D $Z_2$ LGT demonstrates the performance of the ansatz at a large system size. The     infidelity $1-F$ is shown for non-negative momenta in the Brillouin zone in each columns (on a logarithmic scale). Results using the $j^{\rm th}$-order ansatz is shown in the $j^{\rm th}$ row. In general, the ansatz performs better in the strong-coupling regime, which agrees with the results in a smaller system in Sec.~\ref{subsec: ansatz continuum limit numerical analysis}. $F>0.99$ is achieved at the $3^{\rm rd}$ order except for $\ket{k=\frac{6\pi}{13}}$ with $m_f=0.1,0.3$ and $\epsilon=-0.1$. At these couplings, there are other excited states (above the single-particle state) not belonging to the $k=\frac{6\pi}{13}$ sector but with energies very similar to that of $\ket{k=\frac{6\pi}{13}}$, reducing the effectiveness of the 
    MPS ansatz to discern excited states reliably. The MPS states with these coupling are thus less reliable for the fidelity test, and the corresponding parameter range is hashed in the figure.
    }
    \label{fig: scan_26}
\end{figure}

Similar to Sec.~\ref{subsec: ansatz continuum limit numerical analysis}, the fidelity between the order-by-order ansatz states $\ket{k}_{\rm op}$ and the DMRG states $\ket{k}_{\rm MPS}$, now assumed to approximate the exact eigenstates with very high accuracy, is presented for each $k$ value belonging to the Brillouin zone. 
However, one caveat is that the DMRG momentum states $\ket{k}_{\rm MPS}$ for $k \neq 0$ are not accessible due to the energy degeneracy between $\ket{k}_{\rm MPS}$ and $\ket{-k}_{\rm MPS}$. 
The momentum quantum number is not specified during the DMRG sweeps. We assume the output DMRG states with the same energy, labeled $\ket{\psi_1}_{\rm MPS}$ and $\ket{\psi_2}_{\rm MPS}$, are approximately a superposition of the actual momentum eigenstate $\ket{\pm k}_{\rm MPS}$:
\begin{align}
    &\ket{\psi_1}_{\rm MPS} \approx c_1^+ \ket{k}_{\rm MPS} + c_1^- \ket{-k}_{\rm MPS},\nonumber\\
    &\ket{\psi_2}_{\rm MPS} \approx c_2^+ \ket{k}_{\rm MPS} + c_2^- \ket{-k}_{\rm MPS},    
    \label{eq: k_dmrg}
\end{align}
with ${\left|c_1^+\right|}^2 + {\left|c_1^-\right|}^2 \approx {\left|c_2^+\right|}^2 + {\left|c_2^-\right|}^2 \approx 1$. 
Since the orthogonality between the degenerate states holds for the DMRG algorithm (up to numerical precision), i.e., ${}_{\rm MPS}\langle\psi_1|\psi_2\rangle_{\rm MPS} = {}_{\rm MPS}\langle\ k|-k\rangle_{\rm MPS}=0$, it is easily seen that ${\left|c_1^+\right|}^2 + {\left|c_2^+\right|}^2 \approx {\left|c_1^-\right|}^2 + {\left|c_2^-\right|}^2 \approx 1$. 
Furthermore, since the ansatz is built with specified momentum, ${}_{\rm op}\langle k|-k\rangle_{\rm MPS}=0$. 
As a result, one can still access the fidelity of $\ket{k}_{\rm op}$ using $\ket{\psi_{1/2}}_{\rm MPS}$, since
\begin{align}
    {\big|{}_{\rm op}\langle k|{\psi_1}\rangle_{\rm MPS} \big|}^2 + {\big|{}_{\rm op}\langle k|{\psi_2}\rangle_{\rm MPS} \big|}^2   
    & \approx {\left|c_1^+\right|}^2{\big|{}_{\rm op}\langle k|{k}\rangle_{\rm MPS} \big|}^2 + {\left|c_2^+\right|}^2{\big|{}_{\rm op}\langle k|{k}\rangle_{\rm MPS} \big|}^2 \nonumber\\
    &= {\big|{}_{\rm op}\langle k|{k}\rangle_{\rm MPS} \big|}^2=F.
    \label{eq: F_dmrg}
\end{align}
\begin{figure}[t]
    \centering
    \includegraphics[]{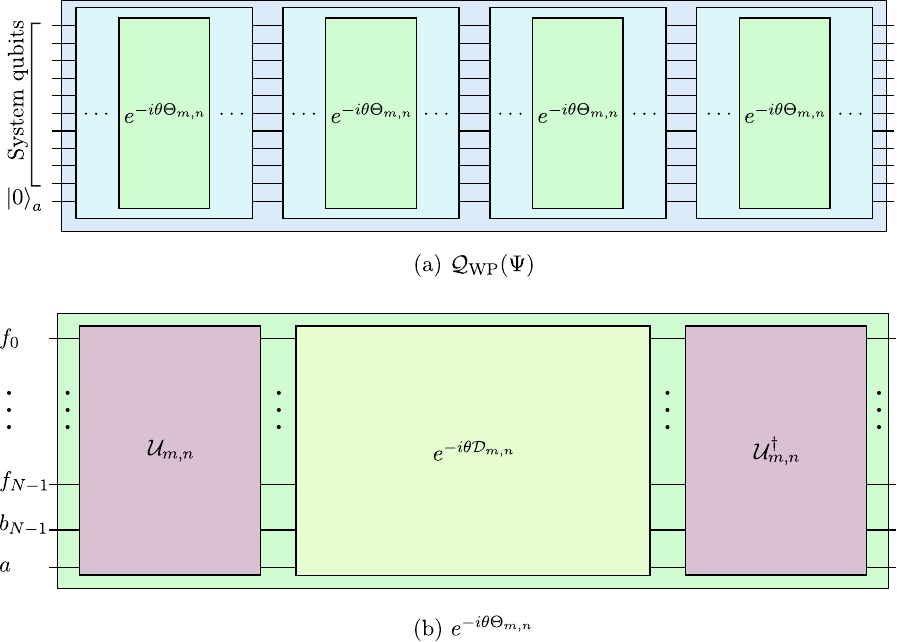}
    \caption{(a) Shown is the circuit decomposition of $\mathcal{Q}_{\rm WP} (\Psi)$ that acts on the system and the ancilla qubits as denoted in Fig.~\ref{fig: Q Init circuit}. Each inner light-blue circuit block represents a Trotter step for the Trotterized angle $\pi/2$ in Eq.~\eqref{eq: ancilla trick}. The dots and the green circuit block within each light-blue block denote that each term after the Trotter expansion is of the form $e^{-i\theta \Theta_{m,n}}$ for $m,n\in \Gamma$, where $\theta$ depends on the Trotter order and Trotter step and $\Theta_{m,n}$ is defined in Eq.~\eqref{eq: Theta mn def}.
    (b) Shown is the circuit for $e^{-i\theta \Theta_{m,n}}$ with $m\neq n$ following Eq.~\eqref{eq: SVD decomposition of exp Theta}. Explicit circuits for the unitary operation $\mathcal{U}_{m,n}$ (pink) and the diagonal operation $e^{-i\theta \mathcal{D}_{m,n}}$ (light green) are presented in Fig.~\ref{fig: example theta circuit} for the choice of $j=1$ in Eq.~\eqref{eq: bk dagger complete ansatz definition}, which we have used in this work.
    Circuits for general $j$ can be constructed similarly.}
    \label{fig: QWP Trotter}
\end{figure}

The fidelity of the ansatz scanning over different $m_f$ and $\epsilon$ values is demonstrated in Fig.~\ref{fig: scan_26}. 
Similar to the parameter scan on a smaller system in Sec.~\ref{subsec: ansatz continuum limit numerical analysis}, only non-negative momenta in the Brillouin zone are plotted for brevity. Similar fidelities are observed for negative-momentum eigenstates. 
Again, the ansatz performs better in the strong-coupling regime, and the order-by-order improvement is visible. Here, the system size is large enough such that the non-mesonic excitations have energy well above that of the $\ket{k=0}$ eigenstate, and thus the cyan contour in Fig.~\ref{fig: scan_10} is not present. 
The $3${\rm rd} order ansatz is able to achieve $F>0.99$, except for the highest momentum $\ket{k=\frac{6\pi}{13}}$ for a small parameter range.
Specifically, for
$m_f=0.1,0.3$ and $ \epsilon=-0.1$, there are excited states in other momentum sectors (above the single-particle eigenstate) with energy very similar to that of $\ket{|k|=\frac{6\pi}{13}}$. It is difficult for the DMRG to discern these from the eigenstates in the $|k|=\frac{6\pi}{13}$ sector.  As a result, the orthogonality originally assumed for the states in Eq.~\eqref{eq: k_dmrg} is not strictly valid, i.e., $\ket{\psi_{1/2}}$ have overlap to states not in the $k=\frac{6\pi}{13}$ sector.
Ideally, a translationally invariant MPS ansatz based on quasiparticle excitations can do a better job at yielding the momentum eigenstates~\cite{haegeman2013elementary,zauner2018variational,haegeman2012variational,Belyansky:2023rgh}.

We conclude that the momentum eigenstates can be faithfully be built from a finite-order ansatz of this work even in larger systems, although for verification purposes, better MPS ansatzes may be needed for large values of momenta to allow a fidelity test. 

\section{Circuit for $\mathcal{Q}_{\rm WP}$ in the minimal-gauge formalism
\label{app: QWP MGF circuit}
}
In this appendix, we describe the circuit for $\mathcal{Q}_{\rm WP}(\Psi)$ that creates the initial single-particle wave packet with wavefunction profile $\Psi$. 
In other words, we aim to implement the operation in Eq.~\eqref{eq: ancilla trick}, where the sum in Eq.~\eqref{eq: Theta Psi def} is exponentiated using the second-order Trotter formula with one Trotter step for the angle $\pi/2$.
Each term in the Trotter expansion is denoted by $e^{-i\theta\Theta_{m,n}}$ for $\theta\in \mathbb{R}$, where $\Theta_{m,n}$ is given by
\begin{equation}
    \Theta_{m,n} \coloneq C_{m,n} \, \widetilde{\mathcal{M}}_{m,n} \otimes |1_{a} \rangle \langle 0_{a}| + C^*_{m,n} \, \widetilde{\mathcal{M}}^\dagger_{m,n} \otimes |0_{a} \rangle \langle 1_{a}|,
    \label{eq: Theta mn def}
\end{equation}
such that,
\begin{equation}
    \Theta_{\Psi} = \sum_{m,n\in \Gamma} \Theta_{m,n}.
    \label{eq: Theta in terms of Theta mn}
\end{equation}
Thus, $\mathcal{Q}_{WP}(\Psi)$ is composed of circuits for $e^{-i\theta\Theta_{m,n}}$, as shown in Fig.~\ref{fig: QWP Trotter}(a).
\begin{figure}[t]
    \centering
    \includegraphics[]{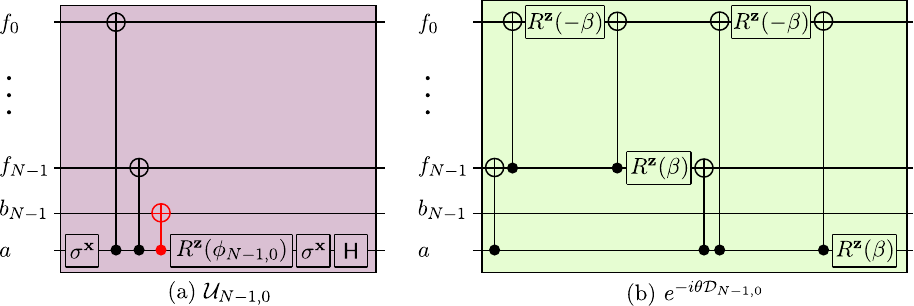}
    \caption{When the ansatz for $b_k^\dagger$ is restricted to have operators smaller or equal to 1-length mesons, i.e., $j=1$ in Eq.~\eqref{eq: bk dagger complete ansatz definition}, the circuits for $\mathcal{U}_{m,n}$ and $e^{-i\theta \mathcal{D}_{m,n}}$ for $|m-n|=1$ can be obtained from the example shown in (a) and (b), respectively. Here, $m=N-1$ and $n=0$ is considered with $C_{N-1,0}=e^{i\phi_{N-1,0}}\left|C_{N-1,0}\right|$ and using a second-order Trotter expansion on the angle $\pi/2$ in Eq.~\eqref{eq: ancilla trick} with one Trotter step. This term includes the gauge-boson qubit $b_{N-1}$ in the MGF. The circuit for $\mathcal{U}_{N-1,0}$ is given in (a), where the system qubits and the ancilla qubit are denoted according the convention explained in the caption of Fig.~\ref{fig: QWP Trotter}. All CNOT gates are controlled on the ancilla qubit with the filled circle denoting the control state $\ket{1}$. The CNOT gates colored in black are targeted on qubits $f_0$ and $f_{N-1}$, while the one in red is targeted on the qubit representing the boson link. The single-qubit gates are the the Hadamard gate $\mathsf{H}$ and rotation gate $R^{\textbf{z}}$ (defined in the caption of Fig.~\ref{fig: GS circuit}). The circuit for $e^{-i\theta \mathcal{D}_{N-1,0}}$ is shown in (b) with $\beta \coloneq \theta \left|C_{N-1,0}\right|/2$ with $\theta=\pi/4$ for a single-step second-order product-formula implementation. The circuits in (a) and (b) remain the same for the operators that do not include the boson link but still satisfy $|m-n|=1$, after making the following changes: 1) adjusting the CNOT-gate positions appropriately, 2) using the corresponding values for $\beta$ and $\phi$, and 3) omitting the red CNOT gate in (a).}
    \label{fig: example theta circuit}
\end{figure}

To implement each $e^{-i\theta\Theta_{m,n}}$, as in our previous work~\cite{Davoudi:2024wyv}, we apply an efficient circuit design based on a singular-value-decomposition (SVD) method~\cite{Davoudi:2022xmb}.
Consider an operator $e^{-i\theta\Theta_{m,n}}$ with $\Theta_{m,n} = A^\dagger\otimes |1_{a} \rangle \langle 0_{a}| + A \otimes |0_{a} \rangle \langle 1_{a}|$ such that $A^2=A^{\dagger 2}=0$ and the SVD decomposition $A=VSW^\dagger$. Then,
\begin{equation}
    e^{-i\theta\Theta_{m,n}} = \mathcal{U}^\dagger e^{-i\theta\mathcal{D}_{m,n}} \mathcal{U},
    \label{eq: SVD decomposition of exp Theta}
\end{equation}
where $\mathcal{U} = \mathsf{H}_a( V^\dagger \otimes |0_{a} \rangle \langle 0_{a}| + W^\dagger \otimes |1_{a} \rangle \langle 1_{a}|) $ and $\mathcal{D}_{m,n} = S \otimes \sigma^\textbf{z}_a$.
The operator $A$ can be read off from Eq.~\eqref{eq: Theta mn def}. It can be easily verified from Table~\ref{tab: bare meson operators} that due to the presence of the $\sigma^{\pm}$ operators in $\widetilde{\mathcal{M}}_{m,n}$ for $m\neq n$, $A^2=A^{\dagger 2}=0$.

The circuit for implementing Eq.~\eqref{eq: SVD decomposition of exp Theta} is shown in Fig.~\ref{fig: QWP Trotter}, with its elements specified in Fig.~\ref{fig: example theta circuit} for the case $j=1$ in Eq.~\eqref{eq: bk dagger complete ansatz definition}, i.e., the bare-meson creation operators $\mathcal{M}_{m,n}$ and their Jordan-Wigner-transformed forms $\widetilde{\mathcal{M}}_{m,n}$ are restricted to create at most 1-length mesons, which we have used in Sec.~\ref{sec: Results}.
In this scenario, $|m-n|\leq 1$ in Eq.~\eqref{eq: Theta mn def}, where $|m-n|$ is defined below Eq.~\eqref{eq: eta j definition}. The circuit for implementing terms with $m=n$ is trivial, and thus, in not discussed here.

Finally, the circuit for $\mathcal{Q}_{\rm WP}(\Psi)$ requires the coefficients $C_{m,n}$.
Recall that, each $C_{m,n}$ depends on the input wavefunction $\Psi(k)$ as well as the ansatz parameters ${\alpha^{(\ell),k}_{0/1}}^*$ which optimize the $b^\dagger_k$ operator. The former component is an input based on the desired scattering state one wants to prepare. The latter are obtained using a VQE method that minimizes the energy of the state prepared by acting $b^\dagger_k$ from Eq.~\eqref{eq: bk dagger complete ansatz definition} on $\ket{\Omega}$.
To obtain the corresponding VQE circuit, note that the $b^\dagger_{k_t}$ operator for a target momentum sector with momentum $k_t$, which is parameterized by $\alpha^{(\ell),k_t}_{0/1}$, is obtained by taking $\Psi(k) = \delta_{k,k_t}$ in Eq.~\eqref{eq: b psi dagger def}.
Thus, the circuit for variational optimization of $b^\dagger_{k_t}$ in Eq.~\eqref{eq: bk dagger complete ansatz definition} is given by the same circuit that prepares the wave packet $\Psi(k)$, i.e., $\mathcal{Q}_{\rm WP}(\Psi)$ with $\Psi(k) = \delta_{k,k_t}$.

\section{Three-wave-packet preparation
\label{app:three-WP}
}
\begin{figure}[t]
    \centering
    \includegraphics[]{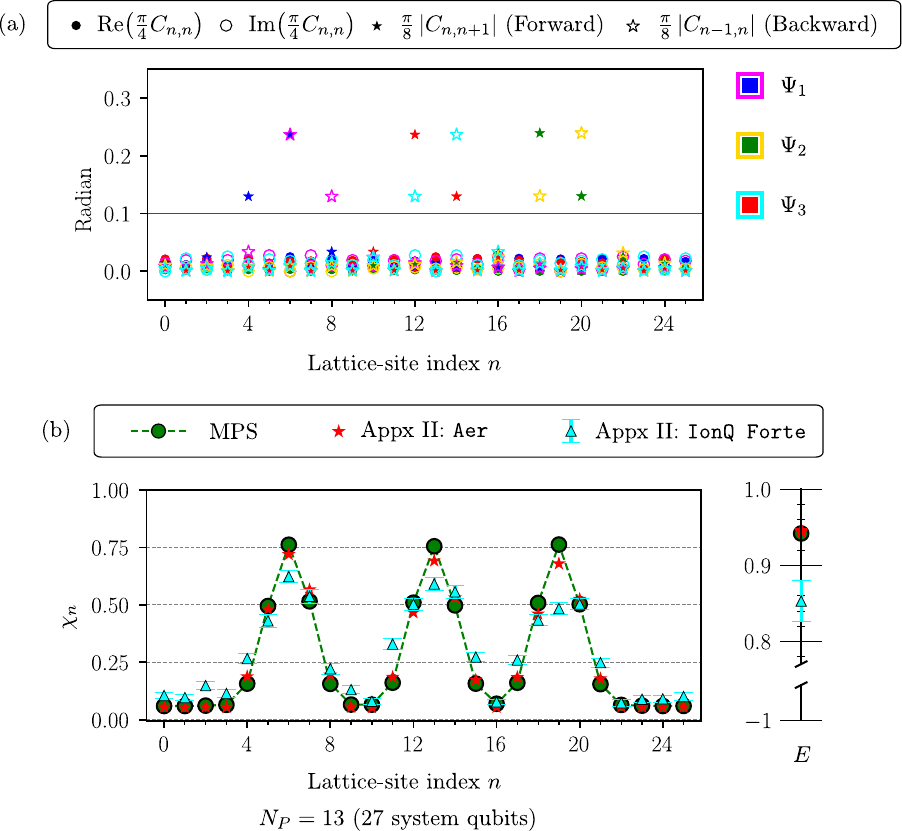}
    \caption{
    (a) Shown are rotation angles for single-qubit gates appearing in the $\mathcal{Q}_{\rm Init}(\Psi_1,\Psi_2,\Psi_2)$ circuit, plotted against their corresponding lattice-site index $n$ for a system of $N_P=13$ physical sites, using Appx II for the preparation circuit. The wave-packet parameters for the first two wave packets are listed in Table~\ref{tab: WP parameters}. The third wave packet is produced with $\mu_3 = 13$, $\sigma_3 = 3\pi/13$ and $\bar{k}_3 = 2\pi/13$ in Eq.~\eqref{eq: Gaussian wavefunction}. The optimized ansatz parameters are provided in Table~\ref{tab: VQE results bk ansatz NP 13}. For other details, see the caption of Fig.~\ref{fig: angle mn plot}.
    (b) Shown are the staggered density $\chi_n$ (left) and the electric field at the boson qubit $E$ (right) for an initial state containing three wave packets on a $N_P=13$ lattice. Only Appx II is used to execute the three-wave-packet version of the $\mathcal{Q}_{\rm Init}$ circuit. The \texttt{Aer} noiseless-simulator results corresponds to $5 \times 10^5$ shots while the \texttt{IonQ Forte} quantum-computer results correspond to 2000 shots.
    \label{fig: 3WP}
    }
\end{figure}
In this appendix, we demonstrate the preparation of an initial scattering state containing three wave packets for an $N_P=13$ system.
Here, the third wave packet is produced with $\mu_3 = 13$, $\sigma_3 = 3\pi/13$ and $\bar{k}_3 = 2\pi/13$ in Eq.~\eqref{eq: Gaussian wavefunction}, and only Appx II is considered for the $\mathcal{Q}_{\rm Init}$ module. The wave packets do not significantly overlap, with the overlap factors, defined in Eq.~\eqref{eq:overlap-def}, being $\left(\Psi_2|\Psi_1\right)= 0.0104$,  $\left(\Psi_3|\Psi_1\right)= -0.0306$, and $\left(\Psi_3|\Psi_2\right)= 0.0059$.
The $\mathcal{Q}_{\rm Init}$ circuit requires 267 CNOT gates, and the circuit was implemented with 2000 shots. The rotation angles, along with the results for the staggered density, are shown in Fig.~\ref{fig: 3WP}.
The hardware results yield $\slashed{Q} = 75.7\,\%$ and $\slashed{a} = 17.70\,\%$ (compared to $\slashed{a}=6.25\,\%$ using the \texttt{Aer} noiseless-simulator with $5 \times 10^5$ shots).
Overall, the three-wave-packet state exhibits the features of the target state obtained from the MPS ansatz for the state.
Lastly, the observable $E$ obtained from the hardware result is in within two standard deviations of the noiseless and the ideal results.

\section{Hadamard test for calculating the return probability
\label{app: Hadamard test}
}
Consider the amplitude
\begin{equation}
    \mathcal{A}(t) \coloneq \braket{\Psi_1,\Psi_2|U(t)|\Psi_1,\Psi_2}.
\end{equation}
The return probability $\mathcal{R}(t)$ is then given by
\begin{align}
    \mathcal{R}(t) \coloneq |\mathcal{A}(t)|^2 = {\rm Re}(\mathcal{A}(t))^2 + {\rm Im}(\mathcal{A}(t))^2,
\end{align}
The real (Re) and imaginary (Im) parts of $\mathcal{A}(t)$ can be computed using a standard Hadamard test.
The state $\ket{\Psi_1,\Psi_2}\otimes\ket{0_{a_t}}$ is first acted with a Hadamard gate on the ancilla $a_t$, followed by a controlled $U(t)$ operation upon the $\ket{1}$ state of the ancilla. 
Finally, a Hadamard or an $R^{\bf x}(\frac{\pi}{4})$ gate is acted on the ancilla to obtain the real or imaginary parts, respectively.
Denote by $p_0$ and $p_1$ the probability for the ancilla to be measured in state $\ket{0_{a_t}}$ and $\ket{1_{a_t}}$, respectively.
Then the quantity $p_0-p_1$ obtains the real or the imaginary part out of the respective circuits.
The circuit for controlled $U(t)$ can be obtained by controlling each block in Fig.~\ref{fig: H epsilon circuit and Q Trott circuit} upon the state of the ancilla at every Trotter step.
The controlled $e^{-i\delta tH^h}$ and $e^{-i\delta tH^m}$ circuits are given by controlling every constituent gate, while the controlled $e^{-i\delta t H^\epsilon}$ is obtained by replacing the $R^{\bf z}$ gates with their controlled versions. This is because in the absence of the $R^{\bf z}$ gates in the circuit in Fig.~\ref{fig: H epsilon circuit and Q Trott circuit}(b), the collective action of the entangling gates is an identity on the state of fermion and the boson qubits.

\section{Time evolution of local observable for $N_P$=13
\label{app: NP13 local observables}
}

\begin{figure}[t]
    \centering
    \includegraphics[]{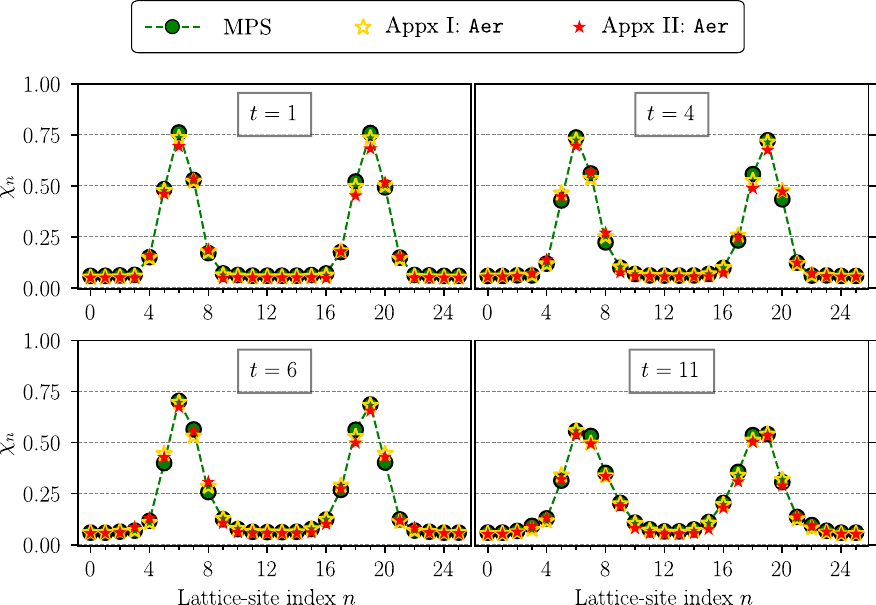}
    \caption{Shown are the staggered density $\chi_n$ corresponding to four instances of time during the evolution, plotted against the lattice-site index $n$.
    The values of $\chi_n$ agree well with the ideal result for both Appxs I and II. These are to be compared with the return-probability results plotted in Fig.~\ref{fig: t evolve return probability}, which exhibit a significant deviation for Appx II.
    \label{fig: NP13 local observable time evolve}
    }
\end{figure}

In Fig.~\ref{fig: t evolve return probability} of the main text, we presented the results for the return probability $\mathcal{R}(t)$ for an $N_P=13$ system obtained from the MPS ansatz, as well as Appxs I and II using the noiseless \texttt{Aer} emulator. We observed significant deviation in this non-local quantity for the cruder Appx II, possibly due to large interference effects. In this appendix, we consider time evolution of the staggered density $\chi_n$ instead, to investigate if the same deviation is seen for a local quantity.

To enable this comparison, we set all the simulation parameters identical to those in Fig.~\ref{fig: t evolve return probability}, i.e., Trotter time step $\delta t=0.25$ with $5\times10^5$ shots for the noiseless emulator.
The results for $\chi_n$ are shown in Fig.~\ref{fig: NP13 local observable time evolve} at four different values of time, $t \in \{1,4,6,11\}$.
These sample values of $t$ are chosen to compare $\chi_n$ when return probability for Appx II shows relatively large deviation from its ideal result (exhibiting 15\% to 30\% relative errors). 
As is observed, this local observable shows very small deviation from the ideal result for both Appxs I and II, confirming that local observables are more robust to small differences in initial states.

\section{Pauli twirling and operator decoherence renormalization
\label{app: Pauli twirling and ODR}}

In this appendix, we describe and implement a noise-mitigation strategy for obtaining the values of $E=\braket{\tilde{\sigma}^{\bf z}}$ under the Trotter time evolution.
The results for this quantity are obtained from runs on the \texttt{IonQ Forte} quantum computer using Appx II, and are displayed in cyan triangles in Fig.~\ref{fig: t evolve E field} and below in Fig.~\ref{fig: t evolve E field ODR}. These results clearly show significant error in this quantity. Even at $t=1$, the value deviates from the ideal result, and it quickly diverges further away from the ideal values during the evolution. On the other hand, as was seen in in Fig.~\ref{fig: inital state 2WP}, there exists reasonable agreement between hardware and simulator results for this quantity at $t=0$ (i.e., after the wave-packets' preparation). We, therefore, conjuncture that the error in the time-evolved observable results from the larger number of entangling gates applied to the boson qubit compared to fermionic qubits, see Fig.~\ref{fig: H epsilon circuit and Q Trott circuit}(b).

The desired result could be recovered, nonetheless, by executing a few additional noise-mitigating circuits, as shown in Fig.~\ref{fig: t evolve E field ODR}, and described below:
\begin{figure}[t!]
    \centering
    \includegraphics[]{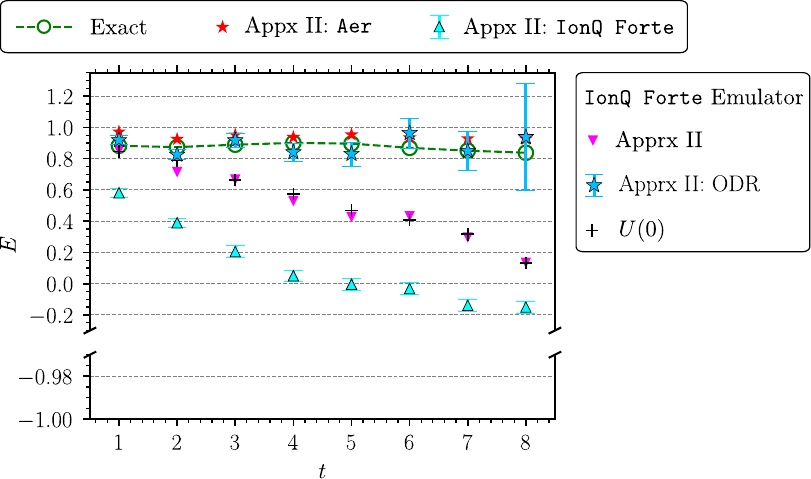}
    \caption{Shown are the noise-mitigated result obtained from Pauli twirling followed by operator decoherence renormalization (ODR) for the electric field at the boson qubit, $E$, after two-wave-packet state evolution for time $t$ in the $N_P$ system (compare with Fig.~\ref{fig: t evolve E field}). The inverted magenta triangles are the results from the Pauli-twirled $U(t)$ circuits. The black crosses are the Pauli-twirled identity circuits, $U(0)$, with similar structure as $U(t)$. Both of these results were obtained using the noisy emulator for the \texttt{IonQ Forte} quantum computer using 3000 shots.
    The error bars obtained from 1000 bootstrap samples are smaller than the markers, and thus, they are not shown here.
    The latter is used to scale the values associated with the inverted magenta triangles to the ODR results, denoted here by blue stars, and the error bars are obtained through error propagation. The results from the \texttt{IonQ Forte} device (cyan stars), which were calculated after removing $\slashed{Q}$ and $\slashed{a}$ errors during post-processing, are also shown for comparison.
    \label{fig: t evolve E field ODR}}
\end{figure}
\begin{itemize}
\item[]
\emph{Pauli twirling:} Using the noisy emulator provided by \texttt{IonQ} for their \texttt{Forte} class devices, we execute 3000 shots for each Trotter step. These are then split over 10 equivalent ``Pauli-twirled'' circuits, and the final result is computed over the aggregated set of shots.
Each Pauli-twirled circuit is obtained by randomly picking one of the following equivalent operations for each CNOT gate in the $e^{-i\delta tH^{\epsilon}}$ block:\footnote{We chose these CNOT gates because the majority of entangling gates acting on the boson qubit belong to $e^{-i\delta tH^{\epsilon}}$.} CNOT, $(\sigma^{\bf z}\otimes \tilde{I})$ CNOT $(\sigma^{\bf z}\otimes \tilde{I})$, $(\sigma^{\bf x}\otimes \tilde{I})$ CNOT $(\sigma^{\bf x}\otimes \tilde{\sigma}^{\bf x})$, and $(\sigma^{\bf x}\otimes \tilde{\sigma}^{\bf y})$ CNOT $(\sigma^{\bf y}\otimes \tilde{\sigma}^{\bf z})$.
Here, tilde operators act on the boson qubit.
This process converts the coherent noise into a decoherent noise. As shown in Fig.~\ref{fig: t evolve E field ODR}, Pauli-twirled results (inverted magenta triangles) improve the results and, in particular, fix the shifted value at $t=1$. This implies that the uniform offset for all time steps likely is caused by coherent errors in the device.
\item[]
\emph{Operator decoherence renormalization (ODR):} Next, we apply a procedure called ODR, which involves scaling the value of the observable $E$ as $E_{\rm ODR}(t) = E(t)/(1-\rho(t))$. The scaling factor $\rho(t)$ is obtained by executing a circuit for each Trotter time $t$ with a similar structure as the target circuit but whose output is known.
We obtain $\rho(t)$ by executing the Pauli-twirled circuits for the $U(0)$ operator, and demand that the value of $E$ remains the same as its value $E_0$ at $t=0$ [shown in Fig.~\ref{fig: inital state 2WP}(a)].
The $U(0)$ circuit can be executed by setting $\delta t \to -\delta t$ in the last two circuit blocks in Fig.~\ref{fig: H epsilon circuit and Q Trott circuit}(a) and using $\epsilon=0$ for the middle block. The $U(0)$ circuits blocks are implemented $n_t$ times, where $n_t$ is the number of Trotter steps up to time $t$. The value of $E'(t) = \braket{\tilde{\sigma}^{\bf z}(t)}$ for these circuits are shown by the black plus sign in Fig.~\ref{fig: t evolve E field ODR}. Note that since the electric-field value at the boson qubit does not change significantly during the time evolution, $E'(t)$ is nearly identical to $E(t)$. The scaling factor $\rho(t)$ is now obtained from these results, and from $E_0$, via the relation $\rho(t) = 1-E'(t)/E_0$. We then apply the same scaling factor $\rho(t)$ to obtain $E_{\rm ODR}$ out of $E(t)$, the results of which are shown as blue stars in Fig.~\ref{fig: t evolve E field ODR}. These results are in much better agreement with the ideal values than unmitigated results, proving the effectiveness of these mitigation strategies in this simulation.
\end{itemize}
This analysis is performed on the \texttt{IonQ Forte} emulator instead of the \texttt{IonQ Forte} quantum system due to the limited hardware access. It would be interesting to see if this procedure also works for the actual hardware data.

\section{Tables of variational-quantum-eigensolver results
\label{app: VQE tables}}
The VQE-optimization results are summarized in this appendix.
The VQE was performed using the noiseless $\texttt{Aer}$ simulator.
Table~\ref{tab: VQE results GS} contains the optimized values of the parameters that characterize $\mathcal{Q}_{\rm GS}$ for $N_P=5$ and $N_P=13$.
Table~\ref{tab: VQE results bk ansatz NP 5} and~\ref{tab: VQE results bk ansatz NP 13} contain the optimization results for the $b^\dagger_k$ ansatz with $j=1$ for $N_P=5$ and $N_P=13$, respectively.

\begin{table}[h!]
    \renewcommand{\arraystretch}{2}
    \begin{center}
        
    \begin{tabular}{|C{1cm}|C{2cm}|C{2cm}|C{2cm}|C{2cm}|C{2cm}|}
    \hline
    $N_P$ & $\theta^{h*}$ & $\theta^{m*}$ & $E^{*}_{\rm GS}$ & $E_{\rm GS}$& $F_{\rm GS}$\\
    \hline
    \hline
    5 & 0.1705 & 0.7846 & -8.8739  & -8.8747 & 0.9998 \\
    \hline
    13 & 0.1707 & 0.7855 & -23.0727  & -23.0743   & 0.9996 \\
    \hline
    \end{tabular}
    \end{center}
    \caption{Shown are the values of $\theta^{h*}$ and $\theta^{m*}$ in Eq.~\eqref{fig: GS circuit} for $N_P=5$ and $N_P=13$, obtained from minimizing the energy of the state $\mathcal{Q}_{\rm GS} (\theta^{h},\theta^{m})\ket{\Omega}_\text{SCV}$ with respect to the Hamiltonian in Eq.~\eqref{eq: Z2 Ham MGF JW in Hh Hm He}. We employ the VQE method and simulate the circuit using a noiseless simulator. $E^{*}_{\rm GS}$ is the energy of the optimized state which agrees, up to sub-percent level, with the exact value $E_{\rm GS}$. The fidelity of the prepared ground state is defined as $F_{\rm GS}\coloneq|\langle{\Omega|\mathcal{Q}_{\rm GS} (\theta^{h*},\theta^{m*})|{\Omega}_{\rm SCV}}\rangle|^2$, where $\ket{\Omega}$ is the target ground state. $E_{\rm GS}$ and $\ket{\Omega}$ for $N_P=5$ were calculated by diagonalizing the Hamiltonian, while for $N_P=13$, they were obtained by finding the lowest-energy state using the DMRG method on an MPS ansatz for the state, see Appendix~\ref{app: tensor network} for details.
    \label{tab: VQE results GS}}
\end{table}
\begin{table}[h!]
    \renewcommand{\arraystretch}{2.7}
    \begin{center}
        
    \begin{tabular}{|C{1.5cm}|C{2cm}|C{2cm}|C{2cm}|C{2cm}|C{2cm}|}
    \hline
     $k$ & ${\alpha^{(1),k}_{0}}^*$ & ${\alpha^{(1),k}_{1}}^*$ & $E^{*}_{k}$ & $E_{\rm k}$ & $F_k$\\
    \hline
    \hline
     $0$ & -0.0957 & 1.1112 & -6.1048 & -6.1260 & 0.9843 \\
     \hline
     $\dfrac{\pi}{5}$ & -3.2695 & -3.0880 & -6.0687  & -6.0825 & 0.9894 \\
     \hline
     $-\dfrac{\pi}{5}$ & -1.0565 & -1.0013 & -6.0687  & -6.0825 & 0.9893 \\
     \hline
     $\dfrac{2\pi}{5}$ & -1.7754 & 1.1020 & -5.9780  & -5.9800 & 0.9982 \\
     \hline
     $-\dfrac{2\pi}{5}$ & -1.5139 & -1.4590 & -5.9779 & -5.9800 & 0.9981 \\
    \hline
    \end{tabular}
    \end{center}
    \caption{Shown are the optimized parameters ${\alpha^{(1),k}_{0/1}}^*$ in Eq.~\eqref{eq: bk dagger complete ansatz definition} for $N_p=5$. Here, $E^*_k$ is the Hamiltonian expectation value of the state $b^\dagger_k\ket{\Omega}$ using the optimized parameters, and $E_k$ is the corresponding target energy associated with the single-particle eigenstate $\ket{k}$ with momentum $k$. The fidelity measure $F_k$ is given by $F_k \coloneq |\langle k|b^\dagger_k|\Omega\rangle|^2$. Both $F_k$ and $E_k$ are computed by diagonalizing the Hamiltonian exactly.\label{tab: VQE results bk ansatz NP 5}}
\end{table}
\begin{table}[h!]
    \renewcommand{\arraystretch}{2.7}
    \begin{center}
        
    \begin{tabular}{|C{1.5cm}|C{2cm}|C{2cm}|C{2cm}|C{2cm}|C{2cm}|}
    \hline
     $k$ & ${\alpha^{(1),k}_{0}}^*$ & ${\alpha^{(1),k}_{1}}^*$ & $E'^{*}_{k}$ & $E'_{\rm k}$ & $F'_k$\\
    \hline
    \hline
     $0$ & -3.1579 & -1.9231 & -20.3043 & -20.3255 & 0.9843 \\
     \hline
     $\dfrac{\pi}{13}$ &-1.9443 & -1.7946 & -20.2987  & \multirow{2}{*}{-20.3187} & \multirow{2}{*}{0.9852}  \\
     $-\dfrac{\pi}{13}$ & -1.9447 & 1.1538 & -20.2986  & & \\
     \hline
     $\dfrac{2\pi}{13}$ & -3.1078 & -2.5162 & -20.2824  & \multirow{2}{*}{-20.2992}& \multirow{2}{*}{0.9876} \\
     $-\dfrac{2\pi}{13}$ & -3.6310 & -3.7909 & -20.2826 &  & \\
     \hline
     $\dfrac{3\pi}{13}$ & -1.2193 & 0.7456 & -20.2567  & \multirow{2}{*}{-20.2688} & \multirow{2}{*}{0.9906} \\
     $-\dfrac{3\pi}{13}$ & -1.7016 & -1.7705 & -20.2570  & & \\
     \hline
     $\dfrac{4\pi}{13}$ & -3.7347 & -3.2568 & -20.2236  & \multirow{2}{*}{-20.2306} & \multirow{2}{*}{0.9945} \\
     $-\dfrac{4\pi}{13}$ & -2.7878 & 0.0941 & -20.2236  &  & \\
     \hline
     $\dfrac{5\pi}{13}$ & -3.8328 & -3.8252 & -20.1854  & \multirow{2}{*}{-20.1882} & \multirow{2}{*}{0.9978} \\
     $-\dfrac{5\pi}{13}$ & -3.3866 & 1.8048 & -20.1856  &  & \\
     \hline
     $\dfrac{6\pi}{13}$ & 1.0874 & 0.5975 & -20.1431  & \multirow{2}{*}{-20.1452} & \multirow{2}{*}{0.9753} \\
     $-\dfrac{6\pi}{13}$ & -1.2578 & -2.7840 & -20.1449  & & \\
    \hline
    \end{tabular}
    \end{center}
    \caption{Shown are the optimized parameters ${\alpha^{(1),k}_{0/1}}^*$ in Eq.~\eqref{eq: bk dagger complete ansatz definition} for $N_p=13$.
    Here, $E'^{*}_k$, $E'_k$ and $F'_k$ have the same definitions as their unprimed counterparts in Table~\ref{tab: VQE results bk ansatz NP 5}, however, they were computed using the DMRG method applied on an MPS ansatz for the states. $E'_k$ are the degenerate energy eigenvalues for the single-particle eigenstates with $k=\pm|k|$. The fidelity measure $F'_k$ is computed using the DMRG states that are degenerate in energy, as described in Appendix~\ref{app: tensor network}.
    \label{tab: VQE results bk ansatz NP 13}}
\end{table}
%

\bibliography{bibi.bib}

\end{document}